\documentclass[a4paper,11pt]{article}
\pdfoutput=1 

\usepackage{jcappub} 

\usepackage[T1]{fontenc} 
\usepackage[utf8]{inputenc}

\renewcommand{\d}{\delta}

\newcommand{\kv}{\mathbf{k}}
\newcommand{\qv}{\mathbf{q}}
\newcommand{\sv}{\mathbf{s}}
\newcommand{\pv}{\mathbf{p}}
\newcommand{\xv}{\mathbf{x}}
\newcommand{\tv}{{\boldsymbol\theta}}

\def\Mpc{h^{-1}\,{\rm Mpc}}

\newcommand{\be}{\begin{equation}}
\newcommand{\ee}{\end{equation}}
\newcommand{\bea}{\begin{eqnarray}}
\newcommand{\eea}{\end{eqnarray}}
\newcommand{\nn}{\nonumber}

\newcommand{\Lu}{\Phi}
\newcommand{\dM}{\delta_{\rm mask}}
\newcommand{\dH}{\delta_{\rm conv}}
\newcommand{\sigsqM}{\sigma^2_{\rm mask}}
\newcommand{\estPmask}{\hat{P}_{\rm mask}}
\newcommand{\estPconv}{\hat{P}_{\rm conv}}
\newcommand{\estPcosmo}{\hat{P}_{\rm cosmo}}
 
\newcommand{\Pobs}{P_{\rm obs}}
\newcommand{\Pmask}{P_{\rm mask}}
\newcommand{\Pconv}{P_{\rm conv}}
\newcommand{\Pcosmo}{P_{\rm cosmo}}
 
\def\pinocchio{\textsc{\scriptsize PINOCCHIO}}

\title{Uncertainty in the visibility mask of a survey and its effects on the clustering of biased tracers}

\author[a]{M. Colavincenzo,}
\author[a,b,d]{P. Monaco,}
\author[c,e]{E. Sefusatti,}
\author[a,b,d]{S. Borgani}

\affiliation[a]{Dipartimento di Fisica, Università di Trieste, Via Tiepolo 11, I-34131 Trieste, Italy}
\affiliation[b]{INAF - Osservatorio Astronomico di Trieste, Via G.B. Tiepolo 11, I-34143 Trieste, Italy}
\affiliation[c]{INAF - Osservatorio Astronomico di Brera, via E. Bianchi 46, 23807 Merate (LC), Italy}
\affiliation[d]{INFN - Sezione di Trieste, via Valerio 2, 34127 Trieste, Italy}
\affiliation[e]{INFN - Sezione di Padova, via Marzolo 8, 35131 Padova, Italy}

\emailAdd{colavincenzo@oats.inaf.it}
\emailAdd{monaco@oats.inaf.it}
\emailAdd{emiliano.sefusatti@brera.inaf.it}
\emailAdd{borgani@oats.inaf.it}

\abstract{The forecasted accuracy of upcoming surveys of large-scale structure
cannot be achieved without a proper quantification of the error
induced by foreground removal (or other systematics like 0-point photometry offset). 
Because these errors are highly correlated on the sky,
their influence is expected to be especially important at very large
scales, at and beyond the first Baryonic Acoustic Oscillation (BAO). In
this work we quantify how the uncertainty in the visibility mask of a
survey, that gives the survey depth in a specific sky area,
influences the measured power spectrum of a sample of tracers of the
density field and its covariance matrix. We start from a very large
set of 10,000 catalogs of dark matter (DM) halos in periodic
cosmological boxes, produced with the {\pinocchio} approximate
method. To make an analytic approach feasible, we assume
luminosity-independent halo bias and an idealized geometry for the
visibility mask, that is constant in square tiles of physical length
$l$; this should be interpreted as the projection, at the observation
redshift, of the angular correlation scale of the foreground
residuals. We find that the power spectrum of these biased tracers can
be expressed as the sum of a cosmological term, a mask term and a term
involving their convolution. The mask and convolution terms scale like
$P\propto l^2\sigma_A^2$, where $\sigma_A^2$ is the variance of the
uncertainty on the visibility mask. With $l=30-100$ Mpc$/h$ and
$\sigma_A=5-20$\%, the mask term can be significant at
$k\sim0.01-0.1\ h/$Mpc, and the convolution term can amount to $\sim
1-10$\% of the total. The influence of mask uncertainty on power
spectrum covariance is more complicated: the coupling of the
convolution term with the other two gives rise to several mixed terms,
that we quantify by difference using the mock catalogs. These are
found to be of the same order of the mask covariance, and to introduce
non-diagonal terms at large scales. As a consequence, the power
spectrum covariance matrix cannot be expressed as the sum of a
cosmological and of a mask term. More realistic settings (realistic
foregrounds, luminosity-dependent bias) make the analytical approach
not feasible, and the problem requires on the one hand usage of
extended sets of mock catalogs, on the other hand detailed knowledge
of the correlations among errors in the visibility masks.  Our results
lie down the theoretical bases to quantify the impact that
uncertainties in the mask calibration have on the derivation of
cosmological constraints from large spectroscopic surveys. }

\begin{document}
\maketitle
\flushbottom

\section{Introduction}
\label{sec:section1}
Observations of the Cosmic Microwave Background (CMB),
e.g. \citep{2015arXiv150201589P,2011ApJS..192...18K}, in combination
with expansion and LSS probes, have allowed cosmology to enter an age
of high precision: presently, CMB observations alone are able to fix
the values of cosmological parameters with per-cent accuracy. The
$\Lambda$CDM model has passed this tough observational test, and is
now considered as the standard model for cosmology, though it is based
on the two unknown ingredients of DM and dark energy. The latter
manifests itself as a cosmological constant; possible deviations of
its equation of state from the simplest $w=-1$ one would give a
crucial hint to its nature.

This calls for accurate low-redshift constraints, based on
measurements of the large-scale structures (LSS) as traced by
observable objects like galaxies.  Upcoming surveys, like \citep[DES,
][]{2016MNRAS.460.1270D}, Dark Energy Spectroscopic Instrument
\citep[DESI, ][]{2011arXiv1106.1706S,2013arXiv1308.0847L}), Extended
Baryon Oscillation Spectroscopic Survey (eBOSS), Large Synoptic Survey
Telescope \citep[LSST, ][]{2009arXiv0912.0201L}, Euclid
\citep{2011arXiv1110.3193L}, Wide-Field Infrared Survey Telescope
\citep[WFIRST, ][]{2012arXiv1208.4012G} and the Square Kilometer Array
(SKA), will measure galaxy clustering with sub-percent accuracy. In
particular, the BAO  feature \citep{2005ApJ...633..560E,2013AJ....145...10D} in the
clustering signal provides a standard ruler that will be used to make
precision measurements of the geometry of the Universe in the redshift
range from $\sim0$ to $\sim2$. The large volumes observable at the
highest redshifts will allow to probe very large, $\sim$Gpc scales
with very low statistical error. This is needed to detect small
effects as for example primordial non-Gaussianity or other effects due
to General Relativity (GR) or deviations from it \citep[e.g.][]{2016arXiv160600180A}.

With very high statistics, the error budget will be dominated by
systematics. On the largest scales, at or beyond the BAO scale, it
will become of fundamental importance to keep under control the effect
of foregrounds, due both to the zodiacal light and to the Milky Way
(galactic extinction and stellar contamination above all).  These will
act by modulating the survey depth on the sky. A similar modulation will be
due to instrumental or survey features, like 0-point offset of
photometric calibration \citep[see for instance the calibration of
  BOSS photometry][]{2008ApJ...674.1217P,2011MNRAS.417.1350R}; in the
following, for simplicity we will refer to foreground removal as the
process that we are addressing, but our approach is equally valid for
these systematics.  Careful characterization of foregrounds will make
it possible to subtract them. However, the residuals from this
subtraction will, in most cases, be highly correlated on the sky, thus
mimicking large-scale structure. So, the error on foreground
subtraction must be properly propagated to correctly assess the error
on parameter estimation.

A great effort has been devoted to understanding the effect of
Galactic foreground and foreground removal for the 21-cm line emission
in the reionization epoch
\citep[e.g.][]{2005ApJ...625..575S,2008MNRAS.389.1319J}, 21-cm
intensity mapping survey at low redshift
\citep[e.g.][]{2014MNRAS.441.3271W}. For the LSS an example of
foreground analysis is given by the BOSS survey: they analyzed the
potential systematic effects on the galaxy observed density
\citep[][]{2012MNRAS.424..564R} finding that the major contributions
come from stellar density and Galactic extinction.

In this work we focus on the issue of how the uncertainty in the
removal of foregrounds (or other similar systematics, as said above) 
propagates to the measurement of clustering
at the two-point level and to its covariance. 
The number of galaxies in an observed sample is given, on
average, by the integral of the galaxy luminosity function from a
luminosity threshold, determined by the survey flux limit, to
infinity. In realistic cases, the flux limit is modulated on the sky
by foregrounds, and these are typically correlated on large angular
scales. A visibility mask will quantify this effect in order to remove
it, but this removal will be done with some uncertainty. This will result
in a modulation of the luminosity threshold, that will propagate to
the number density of observed objects, creating fake large-scale
structure.

To address this issue, we have used the approximate method {\pinocchio}
\citep{2016arXiv160504788M} to run 10,000 simulations of a box of
$1.5\ h^{-1}$ Gpc. We will consider DM halo ``mock'' catalogs at
redshift $z=1$, where it is possible to have observational access to
large volumes and the lower level of non linearities allows
approximate methods to be more accurate. We use DM halos in place of
galaxies as biased tracers of the density field, and their mass in
place of galaxy luminosity. This simplification of the procedure is
acceptable in this context, as long as clustering on very large scales
is considered and halo mass is simply used to implement the effect of
a varying ``luminosity'' threshold. As a note, a one-to-one
correspondence between luminosity and mass is equivalent to a
simplified halo occupation distribution (HOD) model, as we will
comment in section \ref{sec:section3}.

To derive analytic predictions of the clustering of these mock
catalogs, we build a toy model based on the following assumptions. (i)
A mass-independent bias scheme is implemented. DM halos and galaxies
share the property of having a mass- or luminosity-dependent bias, but
this greatly complicates the analytic approach. We implement
mass-independent bias by shuffling halo masses among the objects, as
explained in section~\ref{sec:section3}. A coincise analysis of the
mass-dependent bias case will be outlined in Appendix
\ref{sec:bias}. (ii) We consider an idealized geometry for the
mask. In a plane-parallel approximation, the plane of the sky is
identified with the $x-y$ plane. This is tiled with squares of
physical length $l$, and for each tile the residual of foreground
subtraction is quantified by drawing a random number from a Gaussian
distribution.  No correlation among tiles is considered, so the length
$l$ is to be interpreted as the projection (in a flat sky
approximation), at the observation redshift, of the angular
correlation length of the residual foreground.

Using measurements of the power spectrum on the 10,000 mock catalogs
(with and without imposing a mask), and comparing them with analytic
predictions, we will show that it is possible to fully quantify the
impact of the visibility mask on the power spectrum of biased tracers.
This can be written as the sum of a pure cosmological term, a pure
mask term, and a term involving their convolution.  The same
computation for the covariance matrix of the power spectrum is much
more complicated, because the convolution term gives rise to a long
list of mixed terms that are not easy to compute analytically, even in
this idealized setting. This leads to the conclusion that the
covariance matrix of the power spectrum cannot be simply written as
the sum of a cosmology term and a mask term.

The paper is organized as follows. Section \ref{sec:section2} is
dedicated to the theoretical description of our idealized 
mask model and its effect on the measurement of $P(k)$. We will
analyze all the extra terms in the power spectrum and its covariance
due to the presence of this foreground, including its coupling with
the cosmological signal. In section \ref{sec:section3} we will
describe the cosmological mock samples used to test our mask model and
a description of the technique used to compute the mask power spectrum
and produce masked catalogs. Section \ref{sec:section4} will present
the results obtained by applying the mask to the mock catalogs, under
the assumption of a galaxy bias independent of luminosity. Section
\ref{sec:section5} will give the main conclusions of the paper.
Appendix~\ref{sec:mixedterms} reports a more detailed list of all the
mixed terms of the power spectrum covariance matrix, while in appendix
\ref{sec:bias} we will briefly discuss the case of
luminosity-dependent bias, highlighting a possible path for an
analytic approach.

\section{Power spectrum of biased tracers in the presence of foregrounds}
\label{sec:section2}

In this section we derive some simple analytical expressions
describing the corrections to the power spectrum (and its covariance)
of a galaxy sample defined by a given, nominal luminosity threshold
$L_0$ when some foregrounds induce local variations $\delta L$ 
to the effective threshold that depend on the position on the sky.

\subsection{Luminosity function and galaxy number density}
\label{sec:lumfun}

Let us consider a flux-limited sample of galaxies with luminosity
function $\bar{\Phi}(L)$, and let $\Lu(\xv,\,L)dL$ be the galaxy
number density at the position $\xv$ with luminosity between $L$ and
$L+dL$ so that $\bar{\Phi}(L)\equiv\langle \Lu(\xv,\, L)\rangle$, with
$\langle\dots\rangle$ (and the bar) denoting averages over a very large volume. In
general, $\Lu(\xv,\,L)$ cannot be factorized into the product of a
luminosity-dependent ($\bar\Phi(L)$) and a position-dependent
function; if this were the case, the amplitude of clustering would be
independent of luminosity. This means that $\Lu(\xv,\,L)$ encodes the
information of luminosity-dependent bias.

If our
determination of a galaxy luminosity is not affected by foregrounds,
the galaxy number density of a sample of galaxy characterised by the
lower luminosity threshold $L_0$ will be given by \be
n(\xv;L_0)=\int_{L_0}^\infty dL\,\Lu(\xv,L)\,, \ee while its mean
value will be \be \bar{n}(L_0)=\int_{L_0}^\infty dL\,\bar{\Lu}(L)\,,
\ee We can characterize spatial fluctuations in the number density of
galaxies of luminosity $L$ by means of the galaxy overdensity
$\d_\Lu(\xv,\,L)$ defined by the relation \be
\label{eq:Lu}
\Lu(\xv,\,\,L) = \bar{\Lu}(L)\,[1 + \d_\Lu(\xv,\,L)]\,.
\ee
For a sample of galaxies with luminosity threshold $L_0$ we define instead the overdensity $\d(\xv;\,L_0)$ by means of the relation
\be
n(\xv;\,L_0) = \bar{n}(L_0)\,[1 + \d(\xv;\,L_0)]\,.
\ee
It follows that the two overdensities $\d_\Lu(\xv,\,L)$ and $\d(\xv;\,L_0)$ are related by
\be
\bar{n}(L_0)\,\delta(\xv;\,L_0)=\int_{L_0}^\infty dL\,\bar{\Lu}(L)\,\d_\Lu(\xv,\,L)\,.
\ee

In an observed sample, our measurement of the galaxy luminosity $L$
will be influenced by foregrounds. The two most obvious cases are
galaxy extinction, that will decrease the observed flux, and zodiacal
light, that will increase the sky noise; contamination by field stars
or survey features (e.g., seeing conditions from the ground or solar
aspect ratio from space), or survey features like modulations of 
0-point of photometric calibration  are other examples. 
For a fixed observed flux
limit, the true limiting magnitude, and then the true density, will be
modulated by these foregrounds, or by any \emph{residual} of a
foreground removal procedure. We expect, with some
generality, such residuals to be highly correlated on the sky and we
model here, in a very simple way, how this correlation affects the
measurement of the galaxy power spectrum and its covariance.

We are interested in studying how a modulation of the intrinsic flux
limit propagates to the observed galaxy density and its correlation
functions. To this aim, we assume that the effect of residual
foregrounds consists in changing locally the luminosity threshold
$L_0$ by a quantity $\delta L(\tv)$, where $\tv$ is a
vector that defines the position on the sky. We further assume that
such perturbations to $L_0$ are small, i.e.  $\delta L/L_0 \ll 1$,
and that, as residuals, they have vanishing spatial mean, that is
$\langle \d L(\tv)\rangle=0$, The {\em observed} galaxy number density
of a sample with nominal threshold $L_0$ will then be written as
\begin{align}
n_{\rm obs}(\xv;\,L_0) & =  \int_{L_0+\d L(\tv)}^{\infty}dL\, \Lu(\xv,\,L) \,,
\nonumber\\ & 
= n(\xv;\,L_0)+\delta n(\xv;\,L_0) \,,
  \label{eqn:nobsNEW}
\end{align}

\noindent
where the second contribution on the r.h.s., defined as  

\be
  \d n(\xv;\,L_0)  = \int^{L_0}_{L_0 + \d L(\tv)} dL \,\Lu(\xv,\,L) \,,
  \label{eqn:Mmask}
\ee
represents the  correction due to the mask foreground residuals. 
We notice that:
\begin{align}
\label{eq:emi}
  \d n(\xv;\,L_0) & = \int^{L_0}_{L_0 + \d L(\tv)} dL \,\bar{\Lu}(L)\,[1+\d_\Lu(\xv,\,L)] \nonumber\\
 & = \int^{L_0}_{L_0 + \d L(\tv)} dL \,\bar{\Lu}(L)\nonumber\\
 &  +\int^{L_0}_{L_0 + \d L(\tv)} dL \,\bar{\Lu}(L)\,\d_\Lu(\xv,\,L) \,.
 \end{align}
In the last equation, the first contribution on the r.h.s., describes
the effect of the fluctuations in luminosity $\d L(\tv)$ on the mean
density and has therefore only an angular dependence. The second
contribution, instead, accounts for the effect of the fluctuations in
the threshold on the density perturbation and it is therefore expected
to be subdominant (although not necessarily negligible).

It is important to stress that, even if $\langle \d L(\tv)\rangle=0$,
we cannot expect that the ensemble average of the correction vanishes,
i.e. $\langle\d n(\xv;\,L_0)\rangle = 0$, because of the nonlinear
dependence on $\d L(\tv)$. In particular, the mean of the second
contribution of equation~(\ref{eq:emi}) vanishes due to the fact that
density perturbations at high redshift are expected to be uncorrelated
to any foreground residual, and that $\langle\d_\Lu\rangle=0$ by
definition.  The first term can be Taylor-expanded: \be
\label{eq:taylor}
\int^{L_0}_{L_0 + \d L(\tv)} dL \,\bar{\Lu}(L) \simeq \bar{\Lu}(L_0)\,
\delta L + \frac{1}{2}\frac{d\bar{\Lu}}{dL}(L_0)(\delta L)^2 + \dots
 \ee

\noindent
It is clear that $\langle \d n(\xv;\,L_0) \rangle$ will be non-zero at
the second-order in $\delta L(\tv)$.
For these reasons, in the definition of the {\em observed} galaxy overdensity $\d_{\rm obs}(\xv;\,L_0)$ given by the usual expression 
\be
n_{\rm obs}(\xv;\,L_0) \equiv  \bar{n}_{\rm obs}(L_0)\,\Bigl[1 +\d_{\rm obs}(\xv;\,L_0)\Bigr]\,,
\ee
the mean value $\bar{n}_{\rm obs}(L_0)$ does not equal to the true mean density $\bar{n}(L_0)$.

\subsection{The case of luminosity-independent bias}
\label{sec:problem}

A possible analytical description of galaxy perturbations in the
presence of residual foregrounds consists in Taylor-expanding the
observed number density $n_{\rm obs}(L_0)$ in the threshold
perturbations $\d L(\tv)$. We will present some basic results of such
exercise in Appendix~\ref{sec:bias}. Instead, in what follows we will
make the assumption that the quantity $\Lu(\xv,\, L)$ can be factorized
as the product of a luminosity-dependent and a position-dependent
function.
In terms of equation~\eqref{eq:Lu}:

\be\label{eq:nobias}
\Lu(\xv,\,L) \simeq \bar{\Lu}(L)\,[1 + \d_\Lu(\xv)]\,.
\ee

\noindent
This factorisation is clearly unphysical, as it amounts to neglecting
any dependence of bias on luminosity, but it allows a great
simplification of the calculations.
 
For instance, we have
\be
n(\xv;\,L_0) = \bar{n}(L_0)\,[1 + \d(\xv)]\,,
\ee
with $\delta(\xv)=\d_\Lu(\xv)$. In particular, the corrections due to foregrounds become
\begin{align}
  \d n(\xv;\,L_0) 
  = & \int^{L_0}_{L_0 + \d L(\tv)} dL \,\bar{\Lu}(L)\nonumber\\
   & +\int^{L_0}_{L_0 + \d L(\tv)} dL \,\bar{\Lu}(L)\,\d(\xv) \nonumber\\
  = & [1+\d(\xv)]\,\int^{L_0}_{L_0 + \d L(\tv)} dL \,\bar{\Lu}(L)\,.
 \end{align}
Introducing now the notation 
\be
  \d n_{\rm mask}(\tv;\,L_0) 
   =  \int^{L_0}_{L_0 + \d L(\tv)} dL \,\bar{\Lu}(L) 
  \label{eqn:dnmask}
 \ee
 for the perturbations in the number density exclusively due to foregrounds (and therefore only dependent on the angle $\tv$) we can write the {\em observed} galaxy number density, dropping the explicit dependence on $L_0$, as
\begin{align}
n_{\rm obs}(\xv) &
= [1 + \d(\xv)]\left[\bar{n}+\d n_{\rm mask}(\tv)\right] \,.
  \label{eqn:nobsNEW2}
\end{align}
The observed density contrast $\delta_{\rm obs}(\xv)$, accounting for both cosmological perturbations and foregrounds effect is defined as 
\be
n_{\rm obs}(\xv)\equiv \bar{n}_{\rm obs}\,[1+ \delta_{\rm obs}(\xv)]\,.
\label{eqn:ntot}
\ee
Noting that $\bar{n}_{\rm obs}= \bar{n}+\langle\d n_{\rm mask}(\tv)\rangle$ we have then
\be
 \bar{n}_{\rm obs}\,[1+ \delta_{\rm obs}(\xv)]= \bar{n}_{\rm obs}\,[1+ \delta(\xv)]\,[1+ \dM(\tv)] \,,
\ee
where we introduced the density contrast
\be
\label{eq:deltamask}
\dM(\tv)\equiv \frac{\d n_{\rm mask}(\tv)-\langle \d n_{\rm mask}(\tv)\rangle}{\bar{n}_{\rm obs}}\,.
\ee

\noindent
It is important to stress that $\dM$ can be seen both as a function of
the sky coordinate $\tv$ and as a function of space coordinate $\xv$,
subject to the constraint of being constant along the lines of sight.
Finally we can express the observed galaxy density contrast $\d_{\rm
  obs}(\xv)$ in terms of the actual density contrast $\d(\xv)$ and the
mask-induced relative density corrections $\dM(\xv)$ (expressed as a function of $\xv$) as \be
\delta_{\rm obs}(\xv) = \delta(\xv)-\dM(\xv)- \dM(\xv)\,\delta(\xv)\,
.
  \label{eqn:delta}
\ee 
Adopting the following convention for the Fourier Transform
\be
  \delta(\textbf{k}) = \frac{1}{(2\pi)^3} \int d^3 \textbf{x}
  \ e^{i\textbf{k}\cdot\textbf{x}}\delta(\textbf{x}) \ ,
  \label{eqn:ruleDk}
\ee
the perturbed density contrast, eq. \eqref{eqn:delta}, in Fourier
space will read:
\begin{align}
\delta_{\rm obs}(\textbf{k}) = & \delta(\textbf{k}) - \dM\,(\textbf{k}) 
 - \int  d^3\textbf{q}\,  \delta(\textbf{q})\, \dM (\textbf{k}-\textbf{q}) \nonumber \\
  = &\delta(\textbf{k}) - \dM(\textbf{k}) - \dH(\textbf{k})
  \label{eqn:deltaObsK}
\end{align}
Here we introduced $\dH \equiv \delta\, \otimes\, \dM$ as the convolution of
$\delta(\xv)$ and $\dM(\xv)$.

It is important to stress that, in our simplified model we can write the
observed density contrast $\delta_{obs}$, eq. \eqref{eqn:delta}, as
a function of the independent quantities $\delta$ and $\delta_{mask}$.

\subsection{Power Spectrum} 

The real-space two-point correlation function for the observed
overdensity $\delta_{\rm obs}(\kv)$ can be simply expanded as
\bea
\langle\delta_{\rm obs}(\kv_1)\,\delta_{\rm obs}(\kv_2)\rangle & \!\!\!\!\!\!= \!\!\!\!\!\!&
\langle\delta(\kv_1)\,\delta(\kv_2)\rangle
+\langle\dM(\kv_1)\dM(\kv_2)\rangle+  \nonumber \\
& &\langle \dH(\kv_1)\, \dH(\kv_2) \rangle \,,
\eea
since $\langle \delta \, \dM \rangle=\langle \delta \,\dH \rangle =\langle \dM \, \dH \rangle=0$ 
(they involve averages of either $\delta$ or $\delta_{\rm mask}$). The total, observed power spectrum
$P_{\rm obs}(\kv)$ will therefore be given by:
\be
  P_{\rm obs}(\kv)\, =  \,\Pcosmo(k) \,+\, \Pmask(\kv) \,+\, \Pconv(\kv)\,,
  \label{eqn:PobsEstim}
\ee
where $\Pconv$ is the convolution of the cosmological and mask power
spectra:
\begin{equation}
\Pconv(\textbf{k}) =
\int d^3\textbf{q} \, \Pmask(\textbf{q})\,\Pcosmo(|\textbf{k}-\textbf{q}|) \ .
\label{eqn:conv} 
\end{equation}
This term is of great importance because it couples the cosmological
signal with the noise coming from the mask. Moreover, the integral
generates scale mixing, thus transferring power among different
scales. The procedure presented above is analogous to that of
computing the effect of the variance of the window function of a
survey on the cosmological power spectrum \citep{2013PhRvD..87l3504T}.

Introducing a simple estimator for the power spectrum such as
\be
  \hat{P}(k_i) = \frac{1}{N_{k_i}}\sum_{\textbf{q}\in k_i}\delta_{\textbf{q}}\delta_{-\textbf{q}}
  \label{eqn:PSestimator}
\ee
where $\textbf{q}\in k_i$ denotes a sum over all modes for which
$k=|\mathbf{k}|$ is in the $i$-th bin of size twice the fundamental frequency, $k_f=2\pi/L$, of the box, we can define the power spectrum covariance matrix as
\be
C_{ij} \equiv {\rm cov}[\hat{P}(k_i),\,\hat{P}(k_j)] = \langle \delta \hat{P}(k_i)\delta \hat{P}(k_j)\rangle 
  \label{eqn:Covdef}
\ee
where $\delta \hat{P}(k_i) = \hat{P}(k_i)-\langle \hat{P}(k_i)\rangle$
is the deviation of the $\hat{P}(k)$, measured in a given
realization, from its ensemble average. It is easy to see that the covariance of the observed power
spectrum $P_{\rm obs}$, eq. \eqref{eqn:PobsEstim}, can then be written as
\begin{eqnarray}
C_{ij}^{\rm obs} & \equiv &  {\rm cov}(\hat{P}_{\rm obs}(k_i),\,\hat{P}_{\rm obs}(k_j)) \nn \\
& =  & {\rm cov}[ \estPcosmo(k_i),\, \estPcosmo(k_j) ]+{\rm cov}[ \estPmask(k_i),\, \estPmask(k_j) ]+ C_{ij}^{mixed} \nn \\
& = & C_{ij}^{cosm} +C_{ij}^{mask} + C_{ij}^{mixed}\,,
  \label{eqn:CovPtot}
\end{eqnarray}
i.e., as a sum of the covariance of the cosmological power spectrum,
the covariance of the mask power spectrum $\Pmask(\kv)$ plus a mixed
term accounting for several contributions that can be written as a
function of higher order correlation functions of the density field
and of the mask:
\bea
C_{ij}^{mixed} &=&\langle \estPconv(k_i)\estPconv(k_j)\rangle  - \langle \estPconv(k_i)\rangle \langle \estPconv(k_j)\rangle + \nn \\
 & &\langle \estPcosmo(k_i)\estPconv(k_j)\rangle  - \langle \estPcosmo(k_i)\rangle \langle \estPconv(k_j)\rangle + \notag \\
 & &\langle \estPmask(k_i)\estPconv(k_j)\rangle  - \langle \estPmask(k_i)\rangle \langle \estPconv(k_j)\rangle +\nn \\
& & \langle\estPcosmo(k_i)\hat{G}(k_j)\rangle + \langle \estPmask(k_i)\hat{G}(k_j)\rangle + \nn \\
& & \langle \estPconv(k_i)\hat{G}(k_j)\rangle + \langle\hat{G}(k_i)\hat{G}(k_j)\rangle 
\label{eqn:Cmix}
\eea where $\hat{G}=2\delta_{\qv}\d_{\rm
  mask,{\qv}}-\delta_{\qv}\d_{\rm conv,\qv}+\d_{\rm mask,\qv}\d_{\rm
  conv,\qv}$, with $\qv \in k_i$. Appendix~\ref{sec:mixedterms}
presents a more extended version of this covariance term.

There will be a scale at which perturbations $\delta$ and $\delta_{mask}$ are of the same
order of magnitude, in which case there is no obvious reason why mixing terms should
be small. A full analytical computation of the additional covariance
contribution in $C_{ij}^{mixed}$ is discouraging even in the context
of our simple model. Instead, to quantify the various terms we will
resort to a numerical assessment taking advantage of the large number
of DM halo catalogs produced for this project, to which we will add
the effect of a mask as explained in the next section.

\section{Simulated Catalogs}
\label{sec:section3}

\subsection{Cosmological Catalogs}

As mentioned in the Introduction, our choice is to use DM halos in
place of galaxies as biased tracers. Moreover, we will use the mass
$M$ of the DM halo in place of the galaxy luminosity $L$. In
particular, the nominal mass threshold, i.e. the minimal mass defining
the halo sample (in absence of foregrounds) will be denoted as $M_0$,
corresponding to the $L_0$ of the previous section. This is equivalent
to applying a minimal HOD model \citep{2002PhR...372....1C} with one
galaxy per halo and a linear relation between halo mass and
luminosity. This is known to be unrealistic, but we considered this
approximation proper for the idealised case presented in this paper.

The simulated catalogs we used for all the measurements are DM halo
catalogs obtained with the approximate method {\pinocchio}
\citep[][see \citep{2016arXiv160507752M} for a
  review of approximate methods]{2002MNRAS.331..587M, 2013MNRAS.433.2389M}. This method is
based on (i) generating a linear density field on a grid, as usually
done for the initial conditions of an N-body simulation; (ii)
estimating the time at which each grid point (or particle) collapses,
using a combination of ellipsoidal collapse model and excursion
set theory; (iii) grouping together collapsed particles into
DM halos with an algorithm that mimics their hierarchical
assembly. Displacements of particles (and DM halos) from their initial
positions are computed using Lagrangian Perturbation Theory
\citep[e.g.][]{1995MNRAS.276..115C}. We use the latest version of the
code, V4.1, presented in \citep{2016arXiv160504788M}, where
displacements were computed with LPT up to the third order, resulting
in a sizable improvement of the predicted power spectrum: the
wavenumber at which the prediction of $P(k)$ drops by 10 per cent,
with respect to an N-body simulation run on the same initial
conditions, increases from $k=0.1\ \Mpc$ to $\sim0.3-0.5\ \Mpc$ at
redshift 0 or 1. This lack of accuracy is not relevant for the present
analysis, that is mostly focused on large scales.

We generated 10,000 realizations of a cubic $1500\ \Mpc$ box, sampled
with $1000^3$ particles. This is, to out knowledge, the largest set of
catalogs of DM halo catalogs ever presented.
The cosmological parameters are
$\Omega_m=0.285$, $\Omega_{\Lambda}=0.715$, $\Omega_b=0.044$,
$h=0.695$ and $\sigma_8=0.285$. We used outputs at $z=1$, 
where, as mentioned in the introduction, 
it is possible to have observational access to
large scales and {\pinocchio} is more accurate.
The particle mass is
$M_{\rm p} = 2.67 \times 10^{11}\ h^{-1}M_{\odot}$.

We will consider a halo sample defined by a mass threshold $M_0 =50
\times M_{\rm p}$. With this choice we have approximately $500,000$
halos in each catalog, corresponding to a number density of $1.5\times
10^{-4} \ h^3\, {\rm Mpc}^{-3}$.

\subsection{Implementation of the mask toy model}
\label{sec:MaskModel}

\begin{figure}
  \centering
  \includegraphics[width=0.45\textwidth]{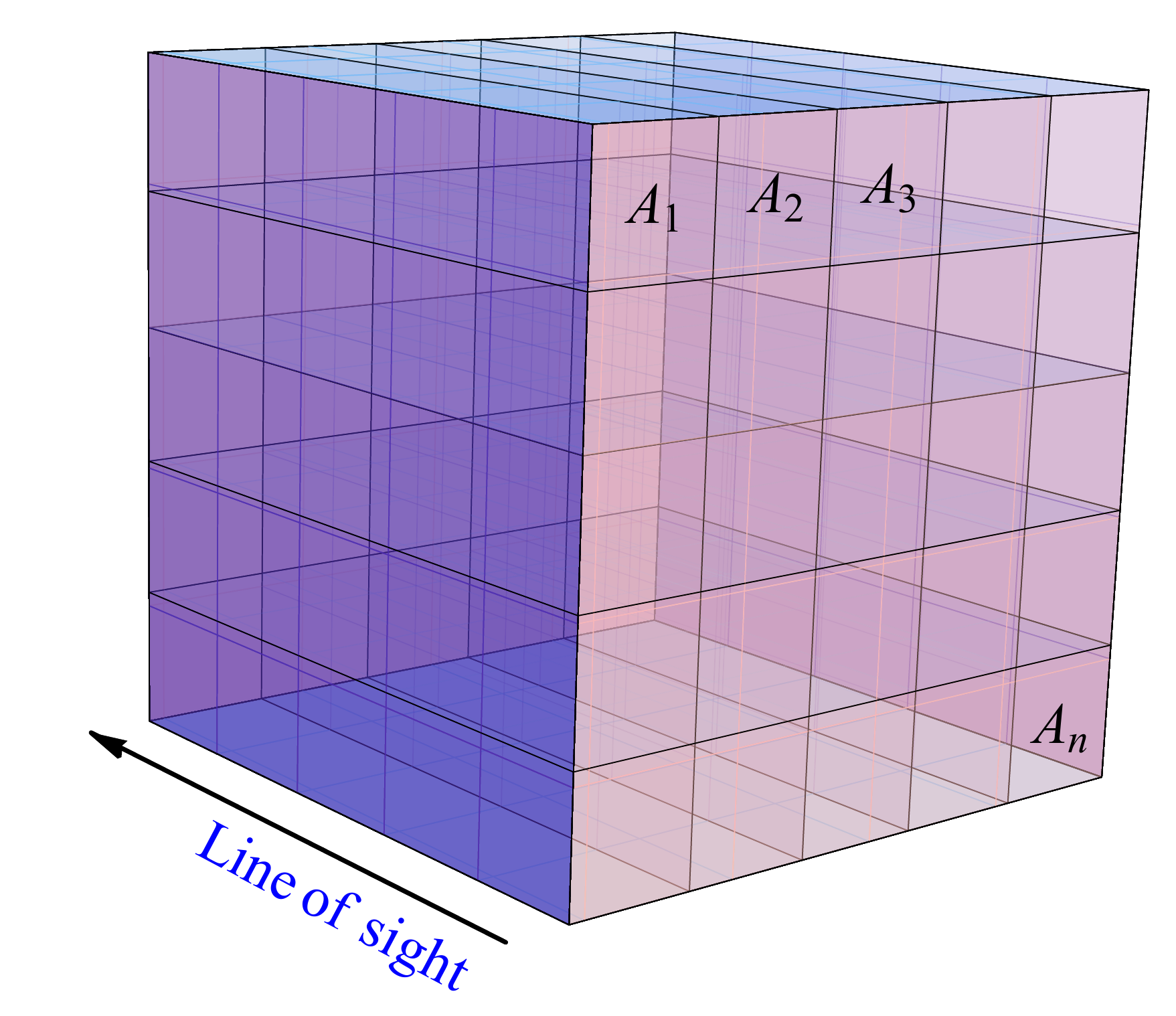}
  \caption{Every simulation box containing the halo catalogs is
    assumed to represent a cosmological volume in the distant-observer
    approximation. Patches of equal foreground error residual are
    modelled as square tiles covering the field-of-view, corresponding
    to one side of the box.}
  \label{fig:box}
\end{figure}

One side of the simulation box will be serving as the field-of-view in
a distant-observer approximation. For simplicity we model patches in
the sky characterised by a constant, uniform foreground residual as
square tiles covering the box side mentioned above. An ``effective''
threshold for halo detection will then be defined, as a correction for
the nominal one $M_0$, for the whole volume (along the line-of-sight)
behind a given tile. Fig.~\ref{fig:box} provides a pictorial
representation of our toy model.

For each halo catalog, we produce a different foreground mask consisting of a correction to the mass threshold $M_0$ for each tile across the field-of-view.  We will describe the relative variation of threshold as the two-dimensional quantity
  \be
  \label{eq:A}
  A(\tv) \equiv \frac{\delta M(\tv)}{M_0} \,. \ee 
Since $A(\tv)$ represents the effect of a residual foreground, we will
assume $\langle A(\tv) \rangle =0$, the bracket representing ensemble
averages.
  
We divide the sky plane into square tiles of length $l$, within which $A$ is kept constant, so we can write:
\be
    A(\tv) = \sum_{i=1}^{N_t}A_i \Theta_i(\tv)
    \label{eqn:maskmodel}
\ee 

\noindent
where the function $\Theta_i(\tv)=1$ if the angular position $\tv$
falls inside the $i$-th tile and zero otherwise, and $N_t$ is the
total number of tiles. The coefficients $A_i$ are assigned as
independent random numbers, drawn from a Gaussian distribution with
standard deviation $\sigma_A$, so $A$-values in nearby tiles are
uncorrelated. The length $l$ therefore represents the physical
correlation length induced by the foreground residual; it will
correspond to the projection, at the observation redshift, of an
angular correlation scale.

The production of masked halo samples proceeds as follows. First, the
DM halo masses provided by {\pinocchio} are modified so as to be
continuous. Indeed, the discreteness due to the particle mass can be
of the same order of the correction to the mass threshold $\delta M$,
leading to spurious effects in the number density that would affect
the covariance matrix. This procedure is applied to all halos with
more than $30$ particles (we altogether ignore smaller groups); this
is smaller than the $50$ particles mass cut mentioned above, because
the mask modulation will decrease the mass cut in $\sim$half of the
sky tiles.  Calling $\alpha$ the logarithmic slope of the DM halo mass
function around $M_0$ (and computing $\alpha$ from the avaraged mass
function of the 10,000 mocks), the halo mass $M$ of a halo made of $N$
particles ($M_{\rm old}=NM_{\rm p}$) is modified as follows:\\
\begin{equation}
  M_{\rm new} = M_{\rm old}
  \left\{1+r\,\left[\left(\frac{N+1}{N}\right)^{\alpha}-1\right]\right\}^{1/\alpha}
  \label{eqn:MassContinue}
\end{equation}
\noindent
where $r$ is a random number between zero and one. Second, in order to
remove the mass dependence of halo bias, that invalidates equation
\eqref{eq:nobias}, in each mock catalog halo masses are randomly
shuffled among all the halos, thus preserving the halo mass function.
In this way, imposing a mass
cut is equivalent to a sparse sampling, and halos with different mass
cuts will have a similar clustering amplitude. Finally, the catalog is
selected by applying the position-dependent mass cut $M_0+\delta
M(\tv)=M_0\,[1+A(\tv)]$.

\subsection{Analytical predictions}

In terms of the adimensional field $A(\tv)$, and adopting now the halo
mass as proxy for the galaxy luminosity, we can rewrite
eq.~\eqref{eq:deltamask}

\bea
\dM(\xv) &  = & \frac{1}{\bar{n}_{obs}}\,\int^{M_0}_{M_0\,[1+A(\tv)]}dM\,\bar{\Phi}(M)-\frac{\langle \delta n_{mask}\rangle}{\bar{n}_{obs}} \nn\\
& = & -\frac{M_0\,\bar{\Phi}(M_0)}{\bar{n}}\,A(\tv)+{\mathcal O}(A^2)-\frac{\langle \delta n_{mask}\rangle}{\bar{n}_{obs}}\,,
\label{eqn:linearN}
\eea

\noindent
showing that the field $A(\tv)$ represents, modulo a $-$ sign, the
overdensity due to the mask, $\dM$, up to a multiplicative constant.

From the definiton of $A(\tv)$, eq.~\eqref{eqn:maskmodel}, it is simple to derive explicitly its Fourier transform
\be
  A_{\kv} = \frac{L_{\rm box} l^2}{(2\pi)^3}\sum_i A_i 
  e^{ik_xx_i}e^{ik_yy_i}
  j_0\left(\frac{k_xl}{2}\right)j_0\left(\frac{k_yl}{2}\right)\delta^K_{k_z,0}
\label{eqn:deltaA}
\ee
where $j_0(x)$ is the zeroth-order Bessel function and $\delta^K$, the
Kronecker symbol. The power spectrum of $A(\tv)$ is given by
\be
  P_A(k_x,k_y,k_z) =
  \frac{L_{\rm box}l^2}{(2\pi)^3}k_f\sigma_A^2j_0^2\left(\frac{k_xl}{2}\right)j_0^2\left(\frac{k_yl}{2}\right)\delta_D(k_z)
  \label{eqn:pkAth}
\ee
where we took the continuum limit by replacing
$\delta^K(\textbf{k})/k_f^3 \rightarrow\delta_D(\textbf{k})$ for $V
\rightarrow \infty$. This term scales as $P_A\propto l^2\sigma_A^2$,
so that at $k<2\pi/l$ it will grow not only, as expected, with the variance of the
residuals but also with the correlation length $l$.
Finally, it is equally simple to write down the convolution of the power spectrum $P_A(\kv)$ with a power spectrum $P(k)$ as
\bea
P_{\rm conv,A}(\kv) & = & \int d^3q \, P_A(\qv)\,\Pcosmo(|\kv-\qv|) \nn\\
 & = & \frac{\sigma_A^2l^2L_{\rm box}k_f}{(2\pi)^2} \int_{k_f}^{k_{max}}
  dq_x \ dq_y \
  j_0^2\left(\frac{q_xl}{2}\right)j_0^2\left(\frac{q_yl}{2}\right) \nn  \\
&  & \times \Pcosmo\left(\sqrt{(k_x-q_x)^2+(k_y-q_y)^2+k_z^2}\right)\,. 
\label{eqn:convA} 
\eea

\noindent
This term, as the previous one, scales as $\propto l^2\sigma_A^2$.
The integral in equation~(\ref{eqn:convA}) can be computed numerically, once $\Pcosmo(k)$ is given. The theoretical prediction for the mask power spectrum makes it
possible to analytically compute the convolution term of eq.~\eqref{eqn:conv}, at least at linear order in $A(\tv)$. 

$P_A(k)$ and $P_{\rm conv,A}(k)$ represent, up to a multiplicative factor, analytical predictions, respectively, for $\Pmask(k)$
 and $\Pconv(k)$, since, as we will see, corrections due to higher
 order terms in $A(\tv)$ are small. Of particular interest is the
 relation between the variance of the mask-induced overdensity,
 $\sigma_{\rm mask}^2\equiv\langle \dM^2\rangle$, and the variance of
 the relative error on the mass threshold,   $\sigma_A^2$. To first
 order in $A$ we have 
\be
  \sigma_{\rm mask}^2 \simeq \frac{M_0^2\bar{\Phi}^2(M_0)}{\bar{n}^2(M_0)}\sigma_A^2\,,
\ee
where $\bar{\Phi}(M_0)$ represents now the halo mass function and $\bar{n}(M_0)$ the number density of objects above the threshold.

\subsection{Power spectrum estimators}
\label{sec:pk}

The power spectrum of the halo catalogs was measured using the
estimator of \cite{2016MNRAS.tmp..928S}, that provides a sophisticated
procedure to minimize the impact of aliasing coming from the estimate
of density of a set of particles on a $600^3$ grid points. 
All the
$k$-bins are multiples of the fundamental frequency of the box, $k_f =
2\pi/L=0.041\ \Mpc$, while the Nyquist frequency is $k_{Nyq}=N_g
k_f/2= 1.256\ \Mpc$, where $N_g=600$ is the grid size.  
The shot noise contribution has always been subtracted.
Given that the
cosmological density field is isotropic in our case, we present here
results for the monopole of the power spectrum; clearly the mask will
induce non-zero multipoles, that will contaminate the redshift space
distorsion signal; we do not address this point in this paper.

\begin{figure}
  \centering
  \includegraphics[width=0.45\textwidth]{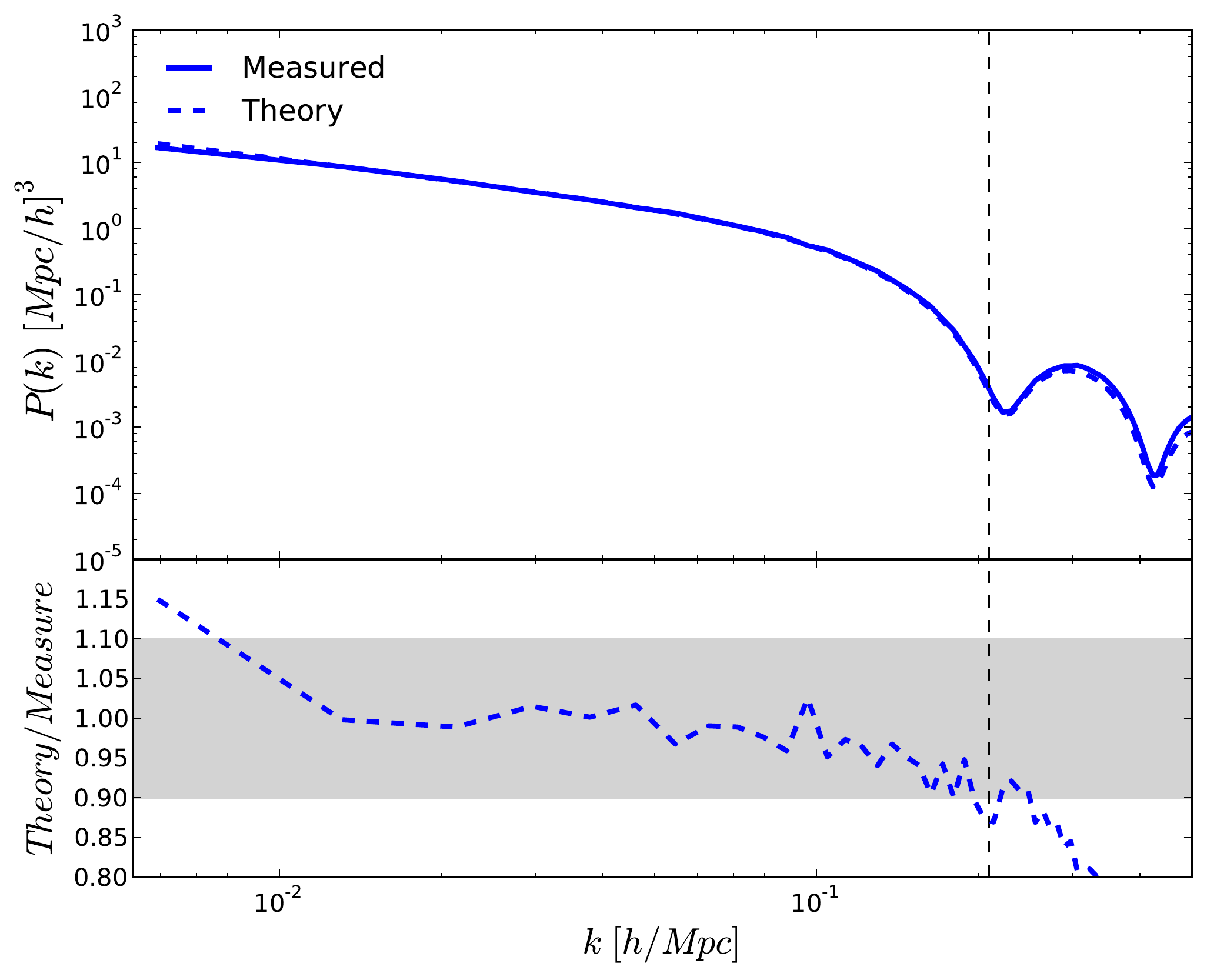}
  \caption{Comparison between measured (continuous line) and analytic
    (dashed line) monopole of the mask power spectrum $P_{\rm
      mask}(k)$. The black dashed vertical line corresponds to
    $k=2\pi/l$. The lower panel gives the residuals.}
  \label{fig:comparing_bin30A20}
\end{figure}

The density field for the estimation of the mask power spectrum,
$\Pmask(k)$, was obtained directly from the two-dimensional field
$A(\tv)$ as follows: $\dM$ is assumed to be equal to the constant
value $A_i$ along the whole i-th tile (fig. \ref{fig:box}), and the
so-defined density field is Fourier-transformed without involving a
density estimate on a set of points. As a consequence, the estimation
of $\Pmask(k)$ is not affected by shot noise. As a consistency test,
we show in figure (\ref{fig:comparing_bin30A20}) the monopole of the
mask power spectrum, computed analytically from eq.~\eqref{eqn:pkAth}
and numerically from 10,000 realisations of the mask $A(\tv)$. The two
results are remarkably consistent at large scales, while at small
scales the numerical determination shows some overestimate with
respect to the analytic one; this is likely due to sampling effects,
but such small differences in the range where the term drops are not a
concern for what follows.

We expect the mask to affect large scales because of its own geometry:
since points behind a given tile are subject to the same effective
threshold $L_0+\delta L$, they will present some level of induced
correlation, even when their separation along the line-of-sight is
very large.

\section{Results}
\label{sec:section4}

The variance of $\dM$, $\sigsqM\equiv \langle\dM^2\rangle$, gives the
magnitude of the effect of the mask on the observed density, and we
are interested in the scale range where it is comparable to the
variance of the density perturbations $\delta$ (of the same order of
the tile length and of the BAO scale).  In fact, the limit $\sigsqM
\ll 1$ corresponds to a very good knowledge of the foregrounds, and
therefore to a negligible effect of possible residuals, while the
opposite limit of large $\sigsqM$ should describe a situation of poor
knowledge of foregrounds that we are expected to avoid.

We will consider the two values 
$\sigma_A= 0.05$ and $0.2$ (or errors of $\sim0.05$ and $\sim0.20$
magnitudes), corresponding respectively to $\sigma_{mask}=0.07$ and
$0.28$. The first value may be a good order of magnitude for a residual foreground (this point will be addressed later), while the second value is very pessimistic and is used to emphasize the effects of foreground removal.
We will also assume two different values for the size of the
tiles, $l=30$ and $100\ \Mpc$. At $z=1$ and for the cosmological
parameters given above, these comoving scales subtend angles of 0.74
and 2.5 degrees. We also tested the case $\sigma_A=0.01$; the effect
of the mask (for this toy case) is entirely negligible for both the
power spectrum and its covariance, so we will not show this case.

\subsection{Power Spectrum}

\begin{figure}
  \centering
  \includegraphics[width=0.45\textwidth]{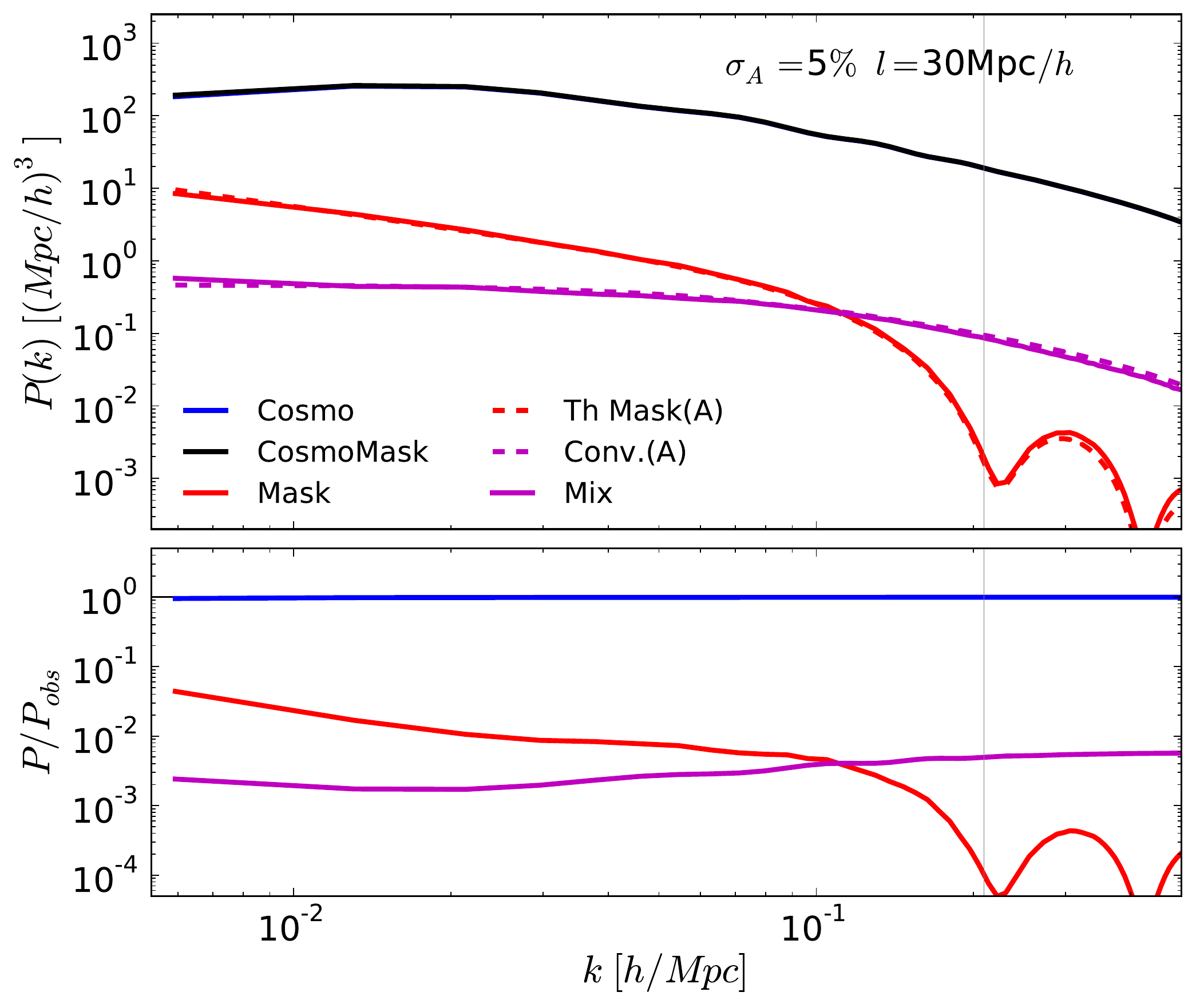}
  \includegraphics[width=0.45\textwidth]{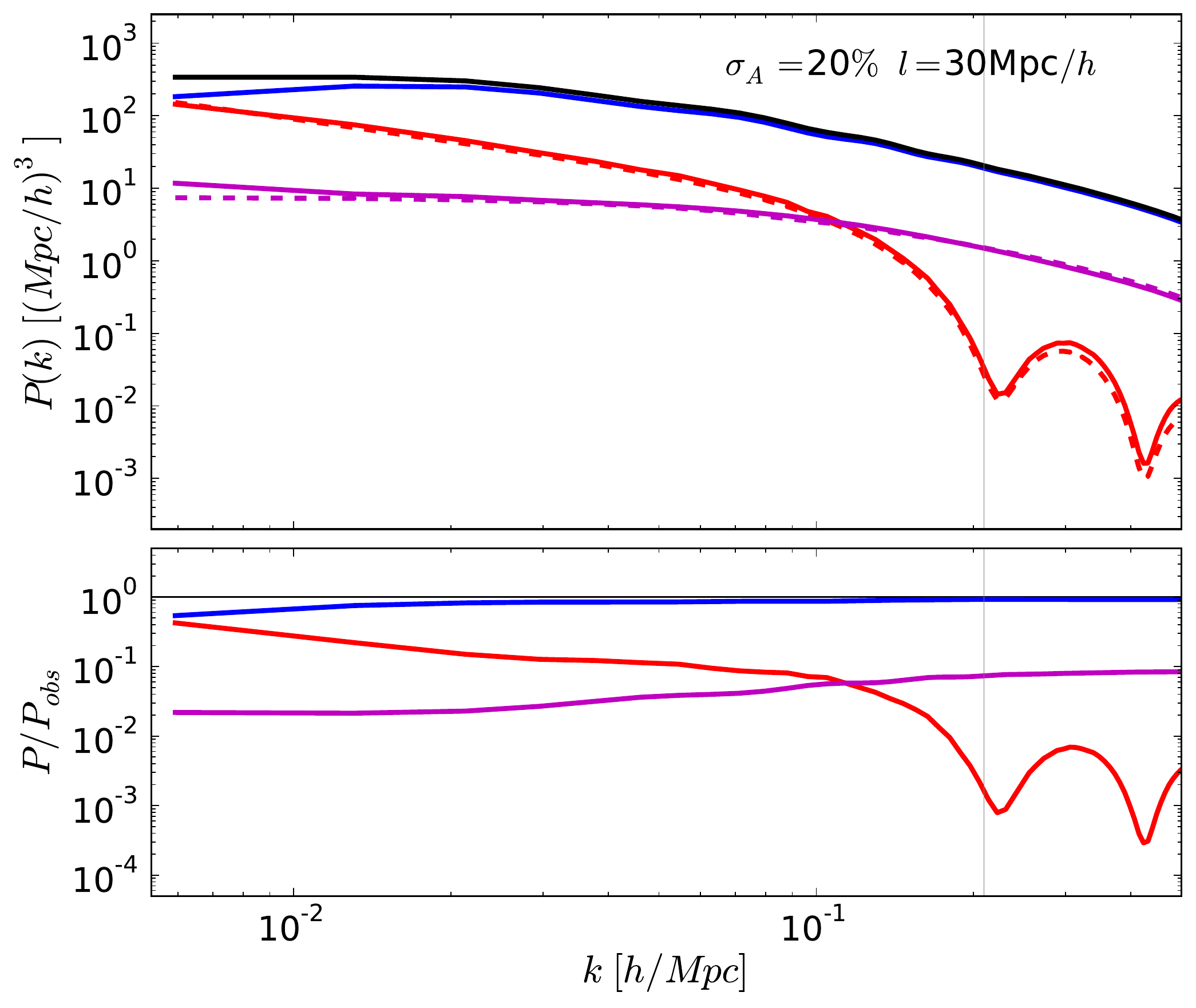}
  \includegraphics[width=0.45\textwidth]{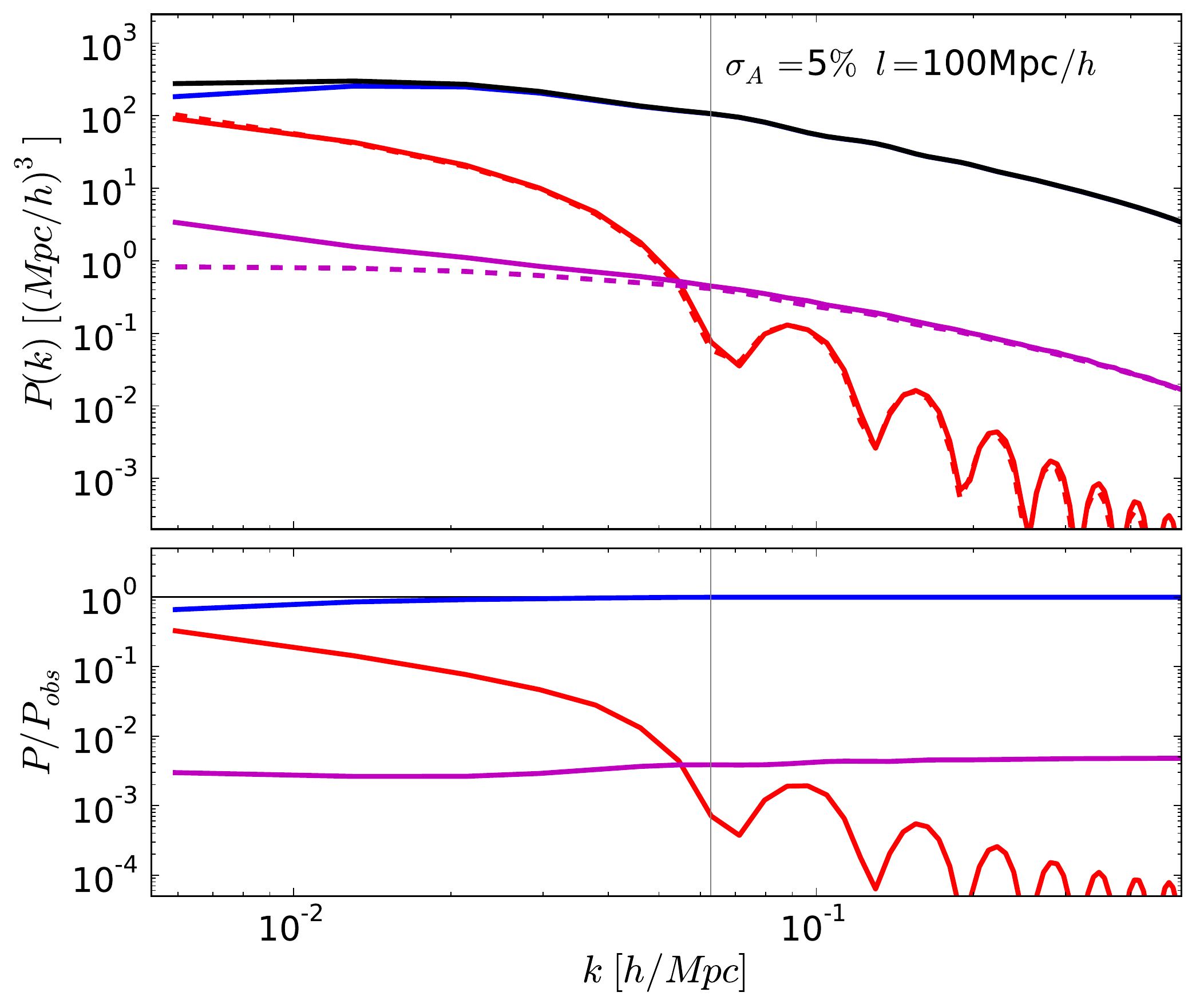}
  \includegraphics[width=0.45\textwidth]{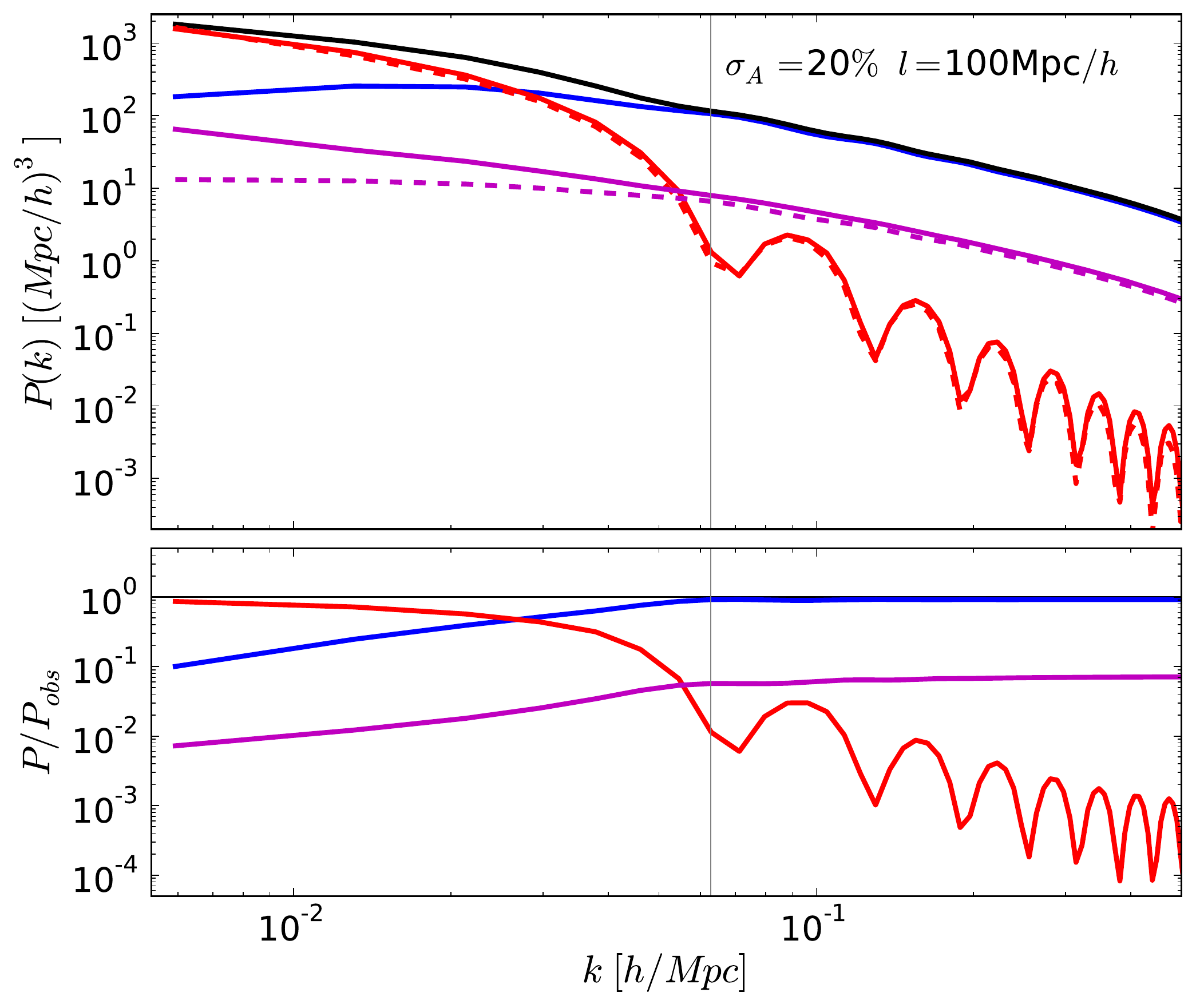}
  \caption{Averaged power spectra of mock catalogs. Top panels: tile size = 30
    $\times$ 30 Mpc$^2/h^2$. Bottom panels: tile size = 100 $\times$
    100 Mpc$^2/h^2$. Left panels: $\sigma_A=5$\%, right panels:
    $\sigma_A=20$\%. In all panels the solid lines denote respectively
    the monopole of total power spectrum (black), cosmological power
    spectrum (blue), mask power spectrum (red) and convolution term
    (magenta). These are all measured from catalogs, the last being
    determined by difference. The dashed red and magenta lines are the
    theoretical predictions for the mask power spectrum (eq.
    \eqref{eqn:pkAth}) and the convolution term (eq.
    \eqref{eqn:convA}). The vertical thin lines mark $k=2\pi/l$. The
    lower plots show the ratio of the various components with respect
    to the total power spectrum; in this case we only show the
    measurements from the mocks.}
  \label{fig:PS_A5A20}
\end{figure}

Figure \ref{fig:PS_A5A20} shows the power spectra for all the four
considered cases, with $l=30$ and $100\ h^{-1}{\rm Mpc} $ (top and
bottom panels) and $\sigma_A=0.05$ and $0.2$ (left and right panels).
Each panel in the figure is composed by two plots. The upper one shows
the monopole of the power spectrum. The coloured lines give the
contributions to the power spectrum of eq. \eqref{eqn:PobsEstim}
(denoted by a black line): the blue line is the cosmological power
spectrum, $\Pcosmo(k)$, the red line represents the pure mask contribution,
$\Pmask(k)$, while the magenta line is the convolution term $P_{\rm
  conv}$ (eq. \eqref{eqn:conv}). Solid lines are obtained by measuring
the 10,000 mocks as explained in Section~\ref{sec:pk}; the convolution
term is computed by difference: 

\be P_{\rm conv} = P_{\rm obs} - P_{\rm cosmo} - P_{\rm mask} \ .
  \label{eqn:Pdiff}
\ee

\noindent
The red and magenta dashed lines represent the theoretical
prediction for the power spectrum of the mask, eq. \eqref{eqn:pkAth},
and for the cross-term, $\Pconv$. The latter is obtained from $P_A(k)$
with the multiplicative factor from eq. \eqref{eqn:linearN}. In the
lower plot of each panel we report the ratio between the components
and the observed power spectrum, to highlight the relative size of
each contribution. In this case we only report the quantities measured
from mocks. Vertical lines in all plots mark $k=2\pi/l$, where the
Fourier transform of the mask starts to oscillate. This is equal to
$\sim0.2\ h/$Mpc for $l=30\ h^{-1}$ Mpc, and to $\sim0.06\ h/$Mpc for
$l=100\ h^{-1}$ Mpc.

As anticipated in Figure~\ref{fig:comparing_bin30A20}, the
contribution of the mask power spectrum (the red line in the plots) is
important at large scales, $k<2\pi/l$. Its relevance depends on
$l\sigma_A$: for $l=30\ h^{-1}$ Mpc it is found to level at about 1
per cent for $\sigma_A=5$\% and in excess of 10 per cent for
$\sigma_A=20$\%, while for $l=100\ h^{-1}$ Mpc its importance gets
increasingly large at large scales even for $\sigma_A=5$\%, while it
dominates at $k<0.03\ h/$Mpc for the higher variance case. At
$k>2\pi/l$ the mask is typically negligible, even though the first
peaks still get above the 1 per cent level in the high variance case.
It is useful to recall that these oscillations are simply the result
of our toy model, and therefore they do not necessarily have a real
physical meaning.

The contribution of the cross-term does not fall down at small scales,
but remains at a fraction of the cosmological power spectrum. At small
scales, this fraction is always below the 1 per cent level for
$\sigma_A=5$\%, but is found to be $\sim10$ per cent in the higher
variance case; notably, this fraction scales with $\sigma_A$ but not
with $l$. The relatively larger contribution of $\Pconv$ with respect
to $\Pmask$ for $k>2\pi/l$ is the result of the transfer of
large-scale power operated by the convolution and does not show the
rather artificial oscillations of the latter. At large scales the
convolution term is always overtaken by $\Pmask$, so it never becomes
dominant, its relative weight ranging from tenths of per cent to few
per cent.  Here we notice a small discrepancy between the theoretical
prediction for $\Pconv$ and the measured one at large scales that
could be due to the small difference between the theoretical
prediction for the mask power spectrum and the measured one at large
scales (fig. \ref{fig:comparing_bin30A20}).

The agreement of analytic and measured contributions allows us to be
confident in the control of the total power spectrum. In the analysis
of the power spectrum covariance we will only use the quantities
measured from the 10,000 mock catalogs.

As a concluding remark, the mask power spectrum $\Pmask(k)$ can easily be
important at large scales even when foreground removal is controlled
to within a few per cent. The reason lies in the scaling with
$(l\sigma_A)^2$: a highly correlated foreground will anyway give a
significant contribution to large scales. Conversely, the convolution
term gives a roughly constant relative contribution to the power
spectrum, that is typically negligible if the uncertainty in the
foreground removal is controlled at the few per cent level, but can
become important in more pessimistic cases. Because the mask creates power on
large scales, within this toy model one could conclude that the BAO
scale should be safe at the per cent level if good control, to the few
per cent level, is achieved on foreground removal. We will get back to
this point in the Conclusions.

\subsection{Covariance}

\begin{figure}
  \centering
  \includegraphics[width=0.45\textwidth]{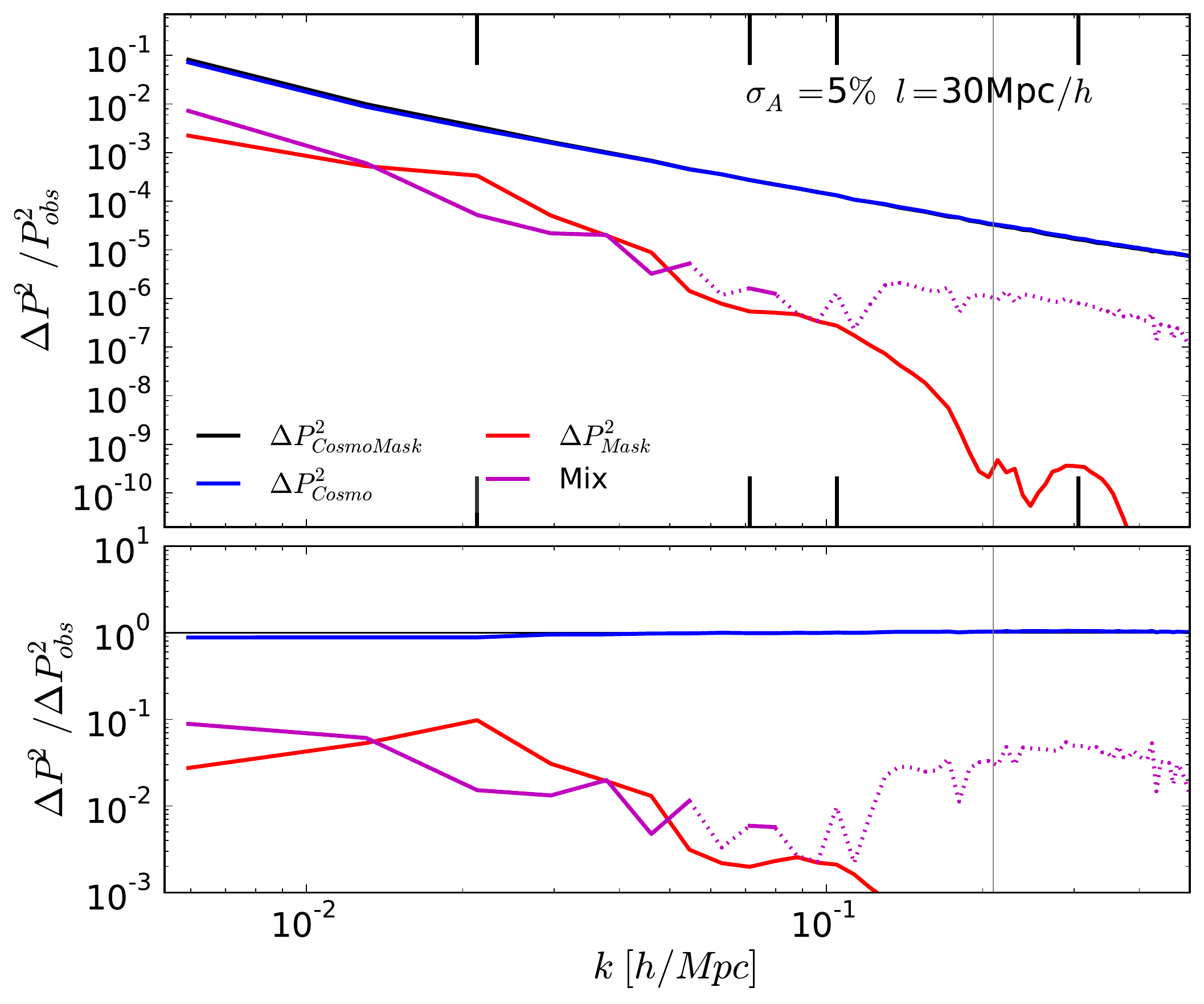}
  \includegraphics[width=0.45\textwidth]{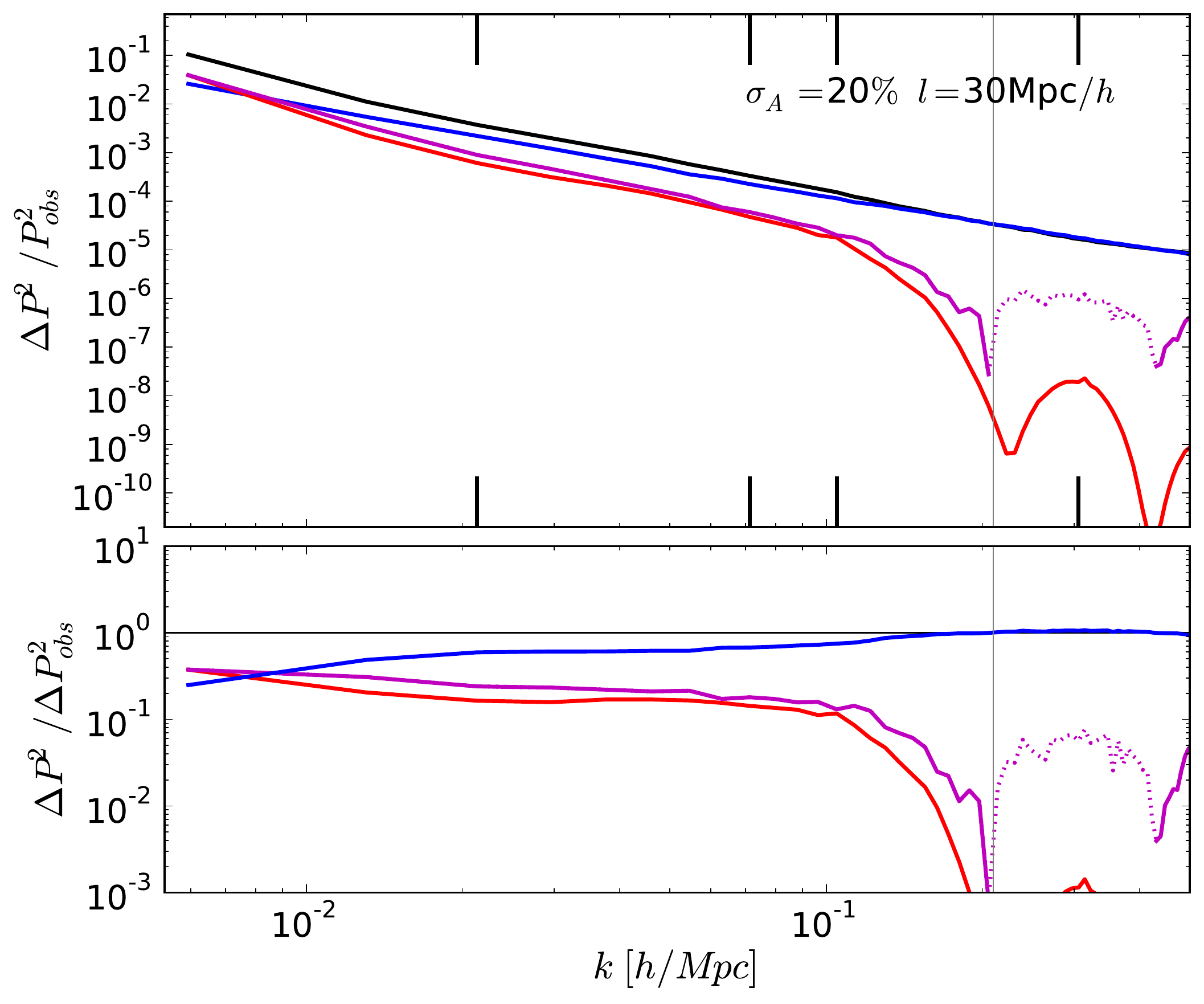}
  \includegraphics[width=0.45\textwidth]{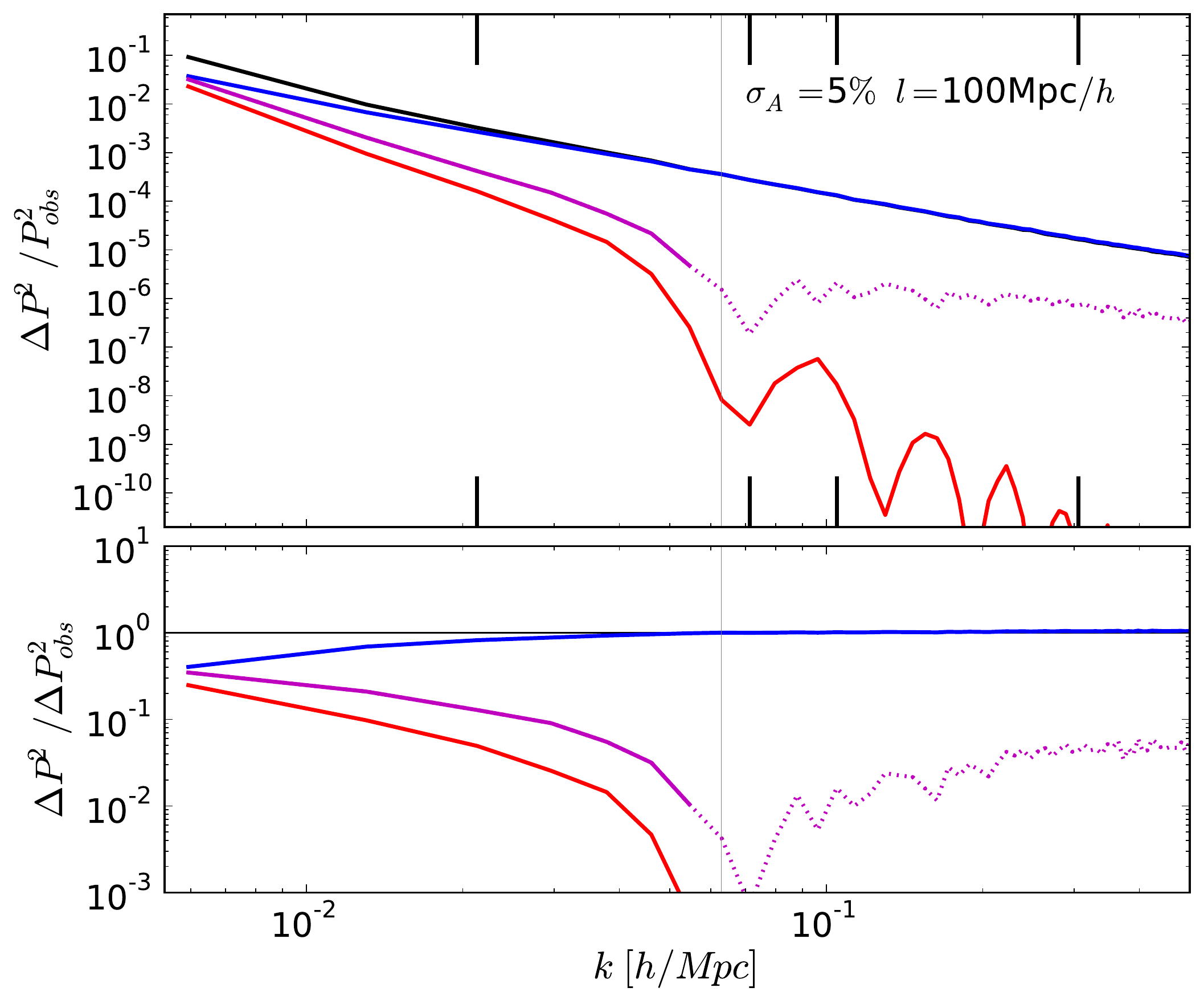}
  \includegraphics[width=0.45\textwidth]{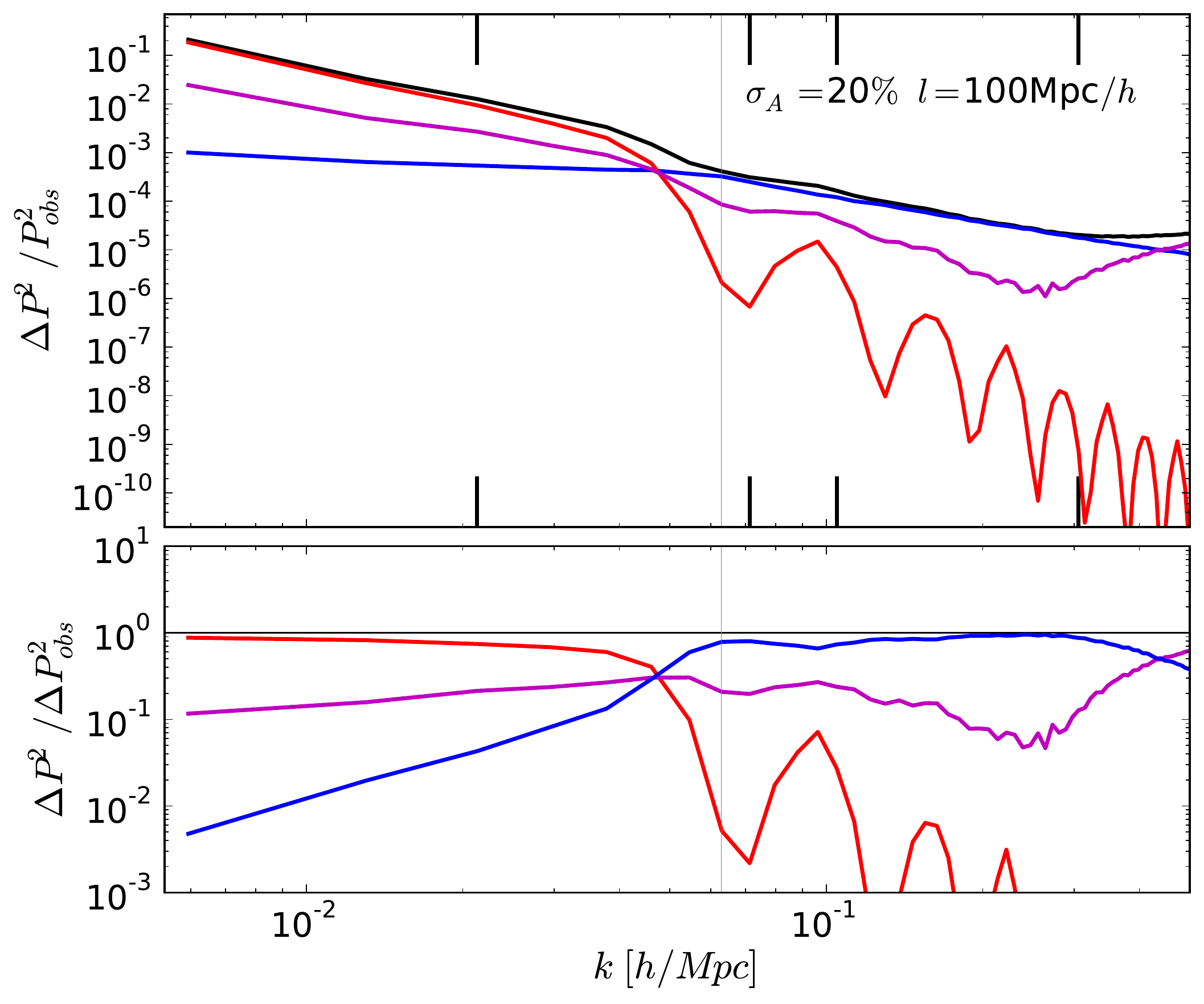}
  \caption{Averaged diagonal components of the covariance matrix of mock
    catalogs. Top panels: tile size = 30 $\times$ 30 $\Mpc$. Bottom
    panels: tile size = 100 $\times$ 100 $\Mpc$. Left panels:
    $\sigma_A=5$\%, right panels: $\sigma_A=20$\%. In all panels the
    solid lines denote respectively the variance of the total power
    spectrum (black), cosmological power spectrum (blue), mask power
    spectrum (red) and all mixed terms (magenta). These are determined
    by difference, dotted lines denote negative values. The vertical
    thin lines mark $k=2\pi/l$. The lower plots show the ratio of the
    various components with respect to the total power spectrum. In
    all figures, the vertical ticks denote the wavenumbers used to
    show the off-diagonal terms in Figure~\ref{fig:corr_coeff_A5A20}.}
  \label{fig:Var_A5A20}
\end{figure}

Figure~\ref{fig:Var_A5A20} shows the variance of the measured power
spectra (all the different components), divided by the measured power
spectrum squared, $\Delta P^2/\Pobs^2$. For a purely cosmological
Gaussian field it would correspond simply to the inverse of the number
of available $k$-modes $1/N_\kv$; in fact, we checked that
our cosmological term is very similar to the Gaussian prediction. 
Here, the magenta curve represents
the mixed mask-cosmology contribution; we recall that $C_{ij}^{\rm
  mixed}$ is not simply the covariance of the convolution $\Pconv$,
but includes a variety of different combinations of cosmological and
error residuals perturbations, in addition to shot-noise. The mixed
term is obtained as the difference:

\be
\Delta P_{\rm mixed}^2 = \Delta \Pobs^2 - \Delta \Pcosmo^2- \Delta \Pmask^2
\,,
  \label{eqn:DPdiff}
\ee

\noindent
since all components on the r.h.s. can be measured independently. The
dotted parts of these curves denote the place where the mixed term
gets negative and then oscillates around zero. In this region the subtraction
of shot noise induces an uncertainty that is larger than the signal
seeked for (the measurement of the mask power spectrum is not affected by
shot noise, so we can detect a much lower signal). The lower plots
show the contribution of each component with respect to the total one,
measured on masked mock catalogs.

Like the power spectrum case, $k=2\pi/l$ marks the scale above which
the pure mask term is important. Comparing the contribution of
different components to the total power spectrum variance we notice
that the pure mask component $\Delta \Pmask^2$ and the mixed one
$\Delta P_{\rm mixed}^2$ present a similar scale-dependence and
comparable amplitudes, at least in the large-scale range where the
mixed terms can be measured.  Starting from the configuration with
$\sigma_A=0.05$ and $l=30\ \Mpc$, the pure mask and mixed terms
contribute to the total variance by a few per cent. Mixed terms are so
small in the first BAO region that they can hardly be measured even
with this statistics. So the contribution of mask terms is modest and
limited to large scales.  Things become pretty different when
$\sigma_A$ and $l$ are increased.  The high variance case gives
contributions of mask and mixed terms well in excess of 10 per cent at
the BAO scale, that become dominant at the largest scales sampled by
the boxes. In the large tile size case the mixed terms get to the 10
per cent level even with the modest mask variance of 5\%, while in the
high variance, large tile size case the covariance is completely
dominated by the mask term on large scales, while the mixed term
remains above the 10 per cent level but still larger than the cosmic
variance. This is the only case where the mixed terms give a
measurable contribution at scales smaller ($k$ higher) than $2\pi/l$;
like the convolution term in the power spectrum, they give a relevant
and non-oscillating small-scale contribution.

\begin{figure}
  \centering
  \includegraphics[width=0.45\textwidth]{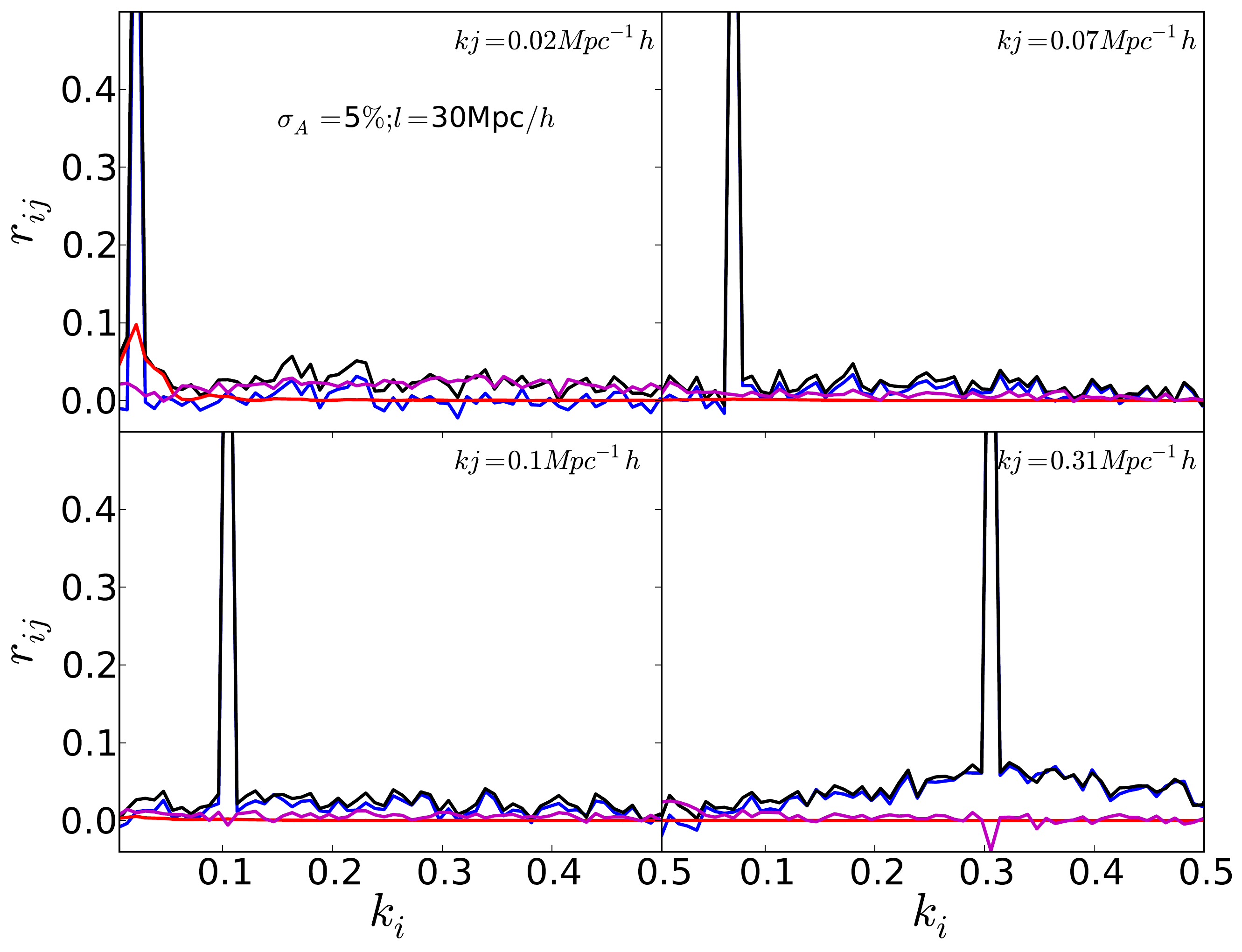}
  \includegraphics[width=0.45\textwidth]{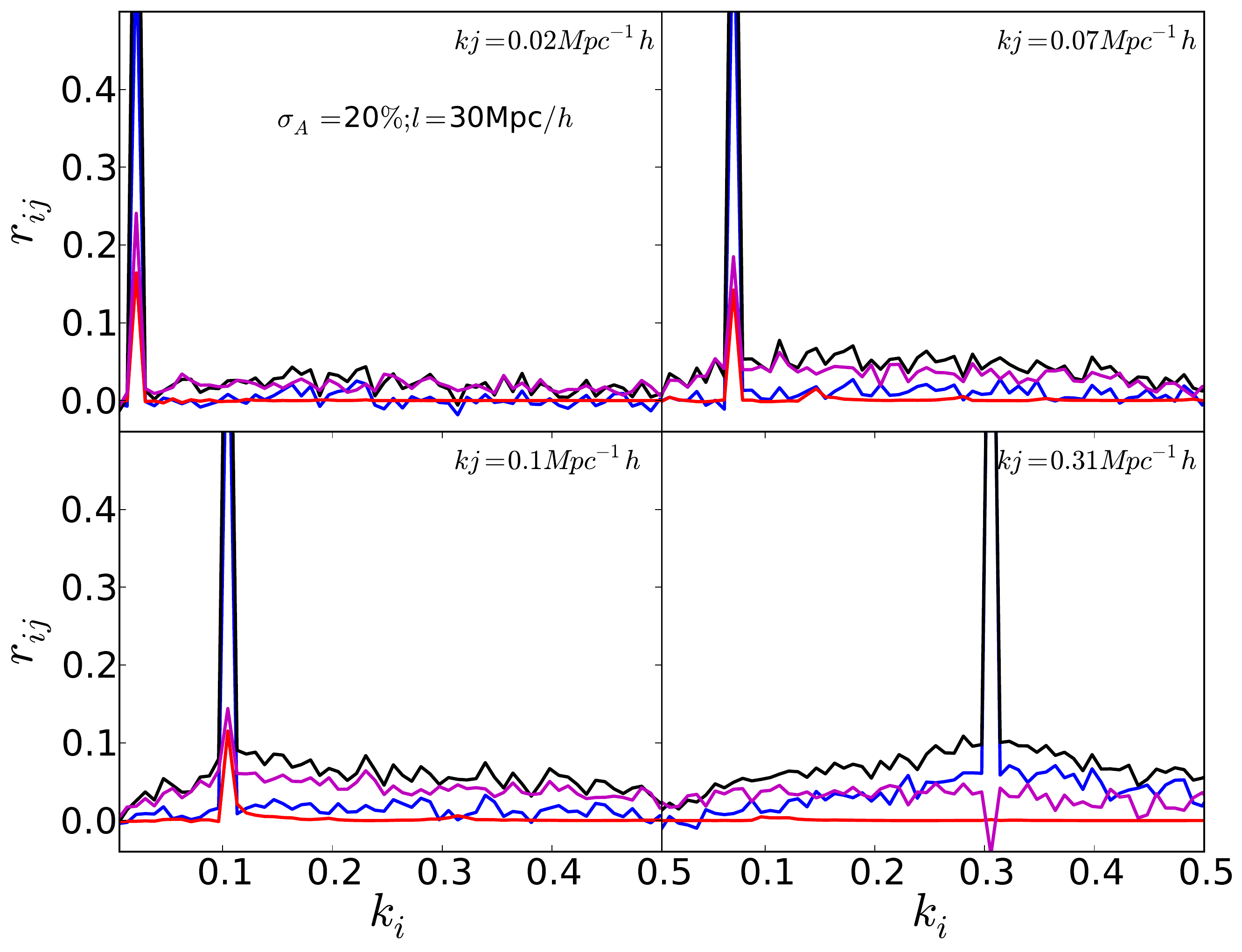}
  \includegraphics[width=0.45\textwidth]{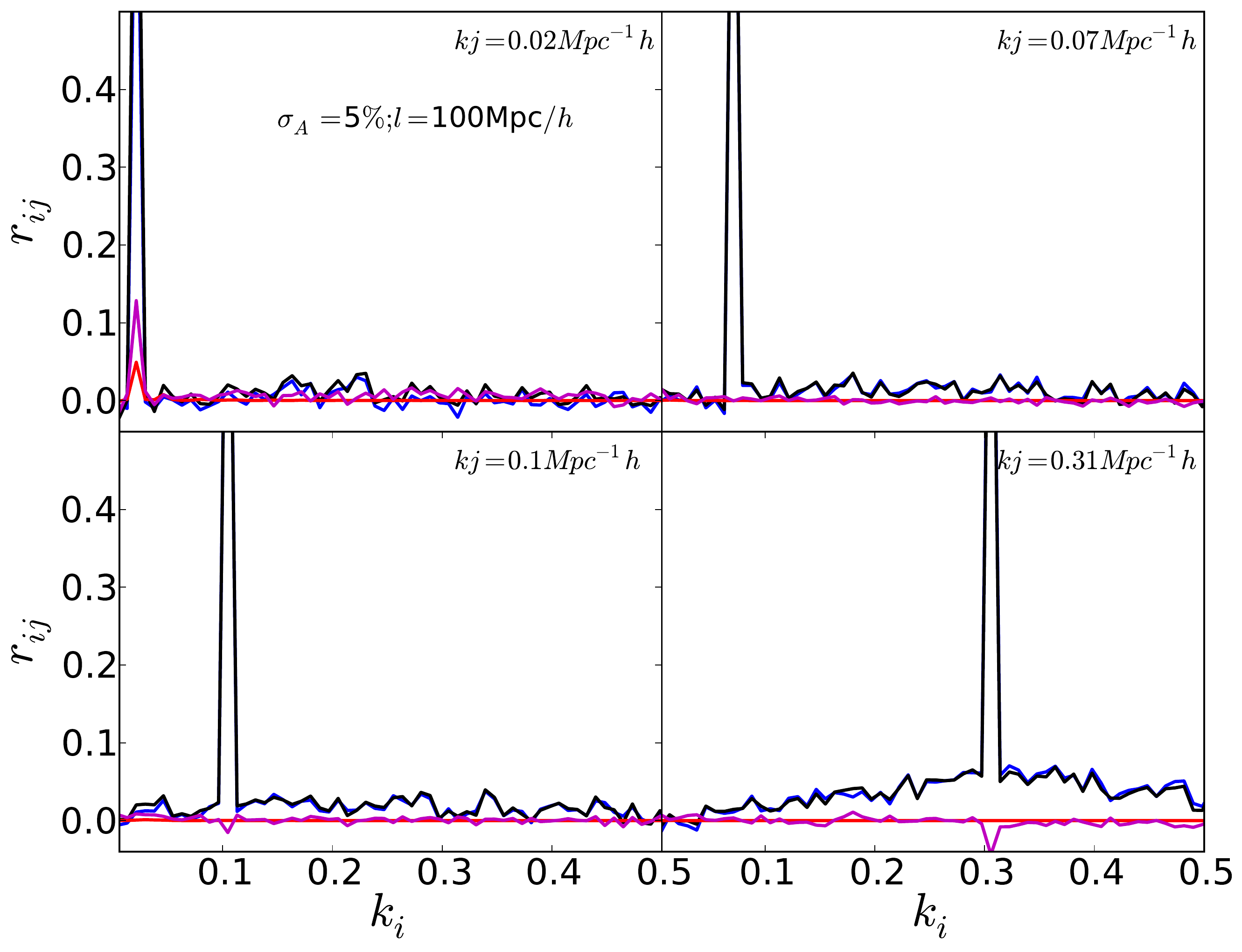}
  \includegraphics[width=0.45\textwidth]{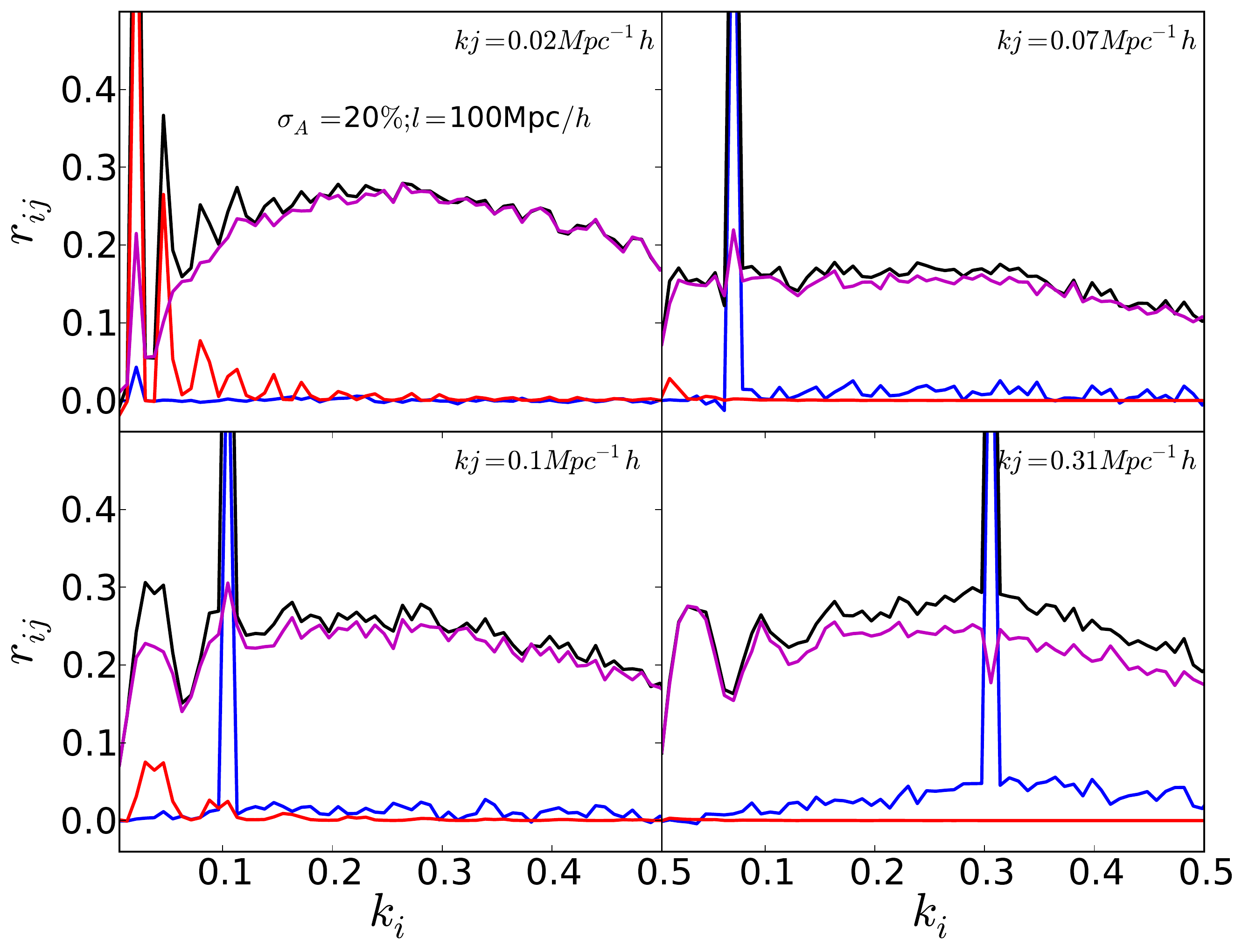}
  \caption{Correlation coefficient: Top panels: tile size = 30 $\times$
    30 $\Mpc$. Bottom panels: tile size = 100 $\times$ 100 $\Mpc$. The
    color code is the same of the previous figures.}
  \label{fig:corr_coeff_A5A20}
\end{figure}

This analysis only shows the diagonal of the covariance matrix.
Off-diagonal terms of the covariance matrix are of great importance,
because they get mixed with the diagonal term during the matrix
inversion necessary to determine the precision matrix that enters the
likelyhood function. We now consider how uncertainties in foreground
subtraction affect the covariance between different wavenumbers by
studying the cross-correlation coefficients defined as:

\be
  r_{ij} = \frac{C_{ij}}{\sqrt{C_{ii}^{obs}C_{jj}^{obs}}} \, .
  \label{eqn:CorrCoeff}
\ee

\noindent
where $C_{ij}$ is defined by eq. \eqref{eqn:Covdef}. These are shows in
figure~\ref{fig:corr_coeff_A5A20} for some relevant values of $k_j$
($0.02$, $0.07$, $0.1$ and $0.31\ h/$Mpc), as a function of
$k_i$. These $k$-values span the range from very large to non-linear
scales, and are marked in Figure~\ref{fig:Var_A5A20} as vertical
ticks.

%

The figure shows how contributions to the normalized covariance due to
mask and mixed terms are negligible in the first configuration with
$\sigma_A=5$\% and $l=30\Mpc$. Off-diagonal terms are small in all
cases, the cosmological one being appreciable at the highest $k_j$.
However, in the smallest $k_j$ bin the mixed terms give a roughly
constant contribution of few per cent. Increasing the scale $l$, we do
not notice a larger impact of mask or mixed terms as we did for the
diagonal; it seems that off-diagonal terms become significant at
$k\ll2\pi/l$, a regime that is not yet reached at $k=0.02\ h/$Mpc in
this case. But when $\sigma_A=0.2$, off-diagonal terms become very
significant, amounting to 5 per cent for $l=30\Mpc$ and are in excess
of 20 per cent for $l=100\Mpc$, independently of scale. In this latter
(rather extreme) case the structure of the power spectrum covariance
matrix is strongly modified; clearly, an inversion of this matrix
without proper account of off-diagonal terms would lead to large
errors in parameter determination.

\section{Conclusions}
\label{sec:section5}

We have addressed the problem of how the uncertainty in the removal of
foregrounds, in an observational survey of biased tracers like
galaxies, propagates to the measurement of the cosmological power
spectrum and its covariance matrix. For this first investigation we
have decided to use a simplified setting, so as to be able to
formulate analytic solutions for the two-point statistics. We have
used DM halos as biased tracers, and their mass as a proxy of galaxy
luminosity, as in a simplified HOD where each halo is populated by a
single galaxy. To this aim, we have produced a very large set of
10,000 realizations of $1.5\ h^{-1}$ Gpc boxes, and extracted DM halos
from these volumes at redshift $z=1$ using the {\pinocchio}
approximate method. This is, to our knowledge, the largest set of
cosmological catalogs of DM halos.  We have neglected luminosity-
(mass-)dependent bias by randomly shuffling masses among DM halos in
each catalog, so as to preserve their mass function. As for the
foreground, we have constructed a simple toy model where, in a
plane-parallel approximation, the $x-y$ plane of the box is tiled in
squares of side $l$, that are characterised by a Gaussian residual
foreground of variance $\sigma_A^2$ that propagates to the density
through a modulation of the mass limit $M_0$; residuals in different
tiles are uncorrelated, so $l$ should be interpreted as the
projection, at the observation redshift, of an angular correlation
scale of the foreground. The chosen values of 30 and 100 $\Mpc$ are
subtended, at $z=1$, by angles of 0.74 and 2.5 degrees respectively.

The main conclusions of our analysis can be summarized as follows:

(1) The residuals of foreground subtraction (``mask'') enter the power
spectrum of masked catalogs as two terms, the power spectrum of the
mask and its convolution with the cosmological power spectrum. This is
similar to what happens when a survey geometry is applied to a
cosmological volume.

(2) The mask term is significant at $k<2\pi/l$, while the convolution
term is usually smaller in this scale range, but can still be
significant at smaller scales due to its scale mixing. Mask and
convolution terms scale as $l^2\sigma_A^2$, implying that large
correlation lengths of the mask residuals may have a significant
effect on large scales even when the foreground removal is controlled
to within a few per cent.

(3) Analytic estimations of mask and mixed power spectrum terms give
results consistent with those measured from mocks, giving
confidence on the level of control of the various terms. 

(4) The power spectrum covariance matrix contains not only the
cosmological and mask contributions, but also several, additional
terms due the coupling of the convolution term with both mask and
cosmology. The sum of all these terms can only be determined by
difference of measurements of masked mocks, cosmological mocks and
pure mask.

(5) Mask and mixed terms are found to have similar effect on the power
spectrum covariance matrix. A 5\% accuracy on foreground removal
guarantees a modest impact of these terms, with the exception of
$k\ll2\pi/l$ modes, where they can significantly contribute to the
diagonal. In this case mixed terms give a roughly constant
contribution to non-diagonal elements of the covariance matrix. The
higher variance case of $\sigma_A=20$\% shows a dramatic impact on the
structure of the covariance matrix.

(6) As a consequence of the relevance of mixed terms, a simple
modeling of the covariance matrix as the sum of a pure cosmological
term and a cosmology-independent term due to the mask appears to be an
oversimplification, as the mixed terms couple cosmology and mask.

As long as BAO is the main target of an observational project, the
results presented in this paper point to the conclusion that a
$\sim5$\% error in foreground removal should guarantee a modest impact
of the mask on parameter estimation. Indeed, due to the
$l^2\sigma_A^2$ scaling, signals with smaller correlation scales will
have little impact, while a large correlation length $l$ will mostly
impact on larger scales (we limited our analysis to correlation
lengths that are subtended, at $z=1$, by relatively small angles
because of the constraints on the box size). These errors can be
compared to estimated errors in foreground removal or photometric
calibration. Clearly, the case $\sigma_A=20$\% is pessimistic, and has
been shown only to illustrate the effect of the mask. Photometric
calibration can be controlled to the millimag level
\citep{2008ApJ...674.1217P}, so its induced errors will likely be
negligible. Conversely, Galaxy extinction is know to the few per cent
level
\citep{1998ApJ...500..525S,2010ApJ...719..415P,2012ApJ...757..166B}
and zodiacal light can have a similar uncertainty far from the
ecliptic \citep{2014A&A...571A..14P}. $\sigma_A=5$\% can then be
considered a realistic order of magnitude for the largest contribution
to the visibility mask uncertainty. However, this conclusion is based
on a very idealized setting, so it should be taken only as an
indication, before tests with much more realistic mocks and masks are
performed. On the one hand, this toy model is mixing modes on the
whole box length; in a realistic survey a redshift bin would span a
smaller comoving distance on the line of sight, and this would reduce
the impact of mixed terms. On the other hand, a more complex mask like
galactic extinction, having power on a range of scales, may easily
have a stronger impact than our toy model; luminosity-dependent bias
would also add to the covariance in a way that needs to be addressed.

To reduce the impact of a foreground to a desired level, one can of
course work to improve the modeling of the foreground and of its
correlated residuals.  But another way to reduce this impact is to
work on the estimator of the two- point statistics, with the aim of
minimizing the impact of the residuals. This has been done, in
preparation to the DESI survey
\citep{2011arXiv1106.1706S,2013arXiv1308.0847L}) by
\citep{Burden_Padmanabhan_et_al2016} for the two-point correlation
function, and by \citep{Pinol_Cahn_et_al2016} for the power
spectrum. In the first paper the authors modify the estimator of the
correlation function to remove the angular mode contaminated by the
incompleteness due to fiber assignment; in the second paper they
investigate different methods to define the survey mean density, in
particulare taking into accunt the fiber assignment coverage.

Two conclusions from the tests we have presented are robust. Firstly,
the impact of foreground removal is of dramatic importance to properly
sample the large scales beyond the BAO. This is expected: foregrounds,
especially the zodiacal light, are correlated on large angular scales,
that are projected to very large scales where the clustering signal is
weak.  But the scaling with $l\sigma_A$ shows that mode coupling gives
a large weight to large-scale correlations, making the control of
residual errors of great importance. It is convenient to recall that
measurement of non-Gaussianity with error on the ${f_{NL}}$ smaller
than unity, or effects of scale-dependent growth related to modified
gravity, should be revealed at scales beyond the power spectrum peak;
therefore the effect of foreground residuals are crucial for these
measurements. Secondly, a poor control of foregrounds can lead to
great changes in the covariance matrix. In particular, a significant
presence of non-diagonal terms has deep consequences in the the
ability to invert the covariance matrix and produce correct
estimations of cosmological parameters and their errorbar. Control of
foregrounds to the few per cent level is confirmed to be of paramount
importance for large-scale structure.

\appendix

\section{Covariance Mix Terms}
\label{sec:mixedterms}
Eq. \eqref{eqn:Cmix} shows all the terms that come from the coupling
of the cosmological signal with the mask. The mixed terms
\bea
C_{ij}^{obs} \supset & & \langle\estPcosmo(k_i)\hat{G}(k_j)\rangle + \langle \estPmask(k_i)\hat{G}(k_j)\rangle + \nn \\
& & \langle \estPconv(k_i)\hat{G}(k_j)\rangle + \langle\hat{G}(k_i)\hat{G}(k_j)\rangle 
\label{eqn:Cmix2a}
\eea
are written in implicit form. We recall that 
\be
\hat{G}=\delta_{\qv}\d_{\rm mask,{\qv}}-\delta_{\qv}\d_{\rm conv,\qv}+\d_{\rm mask,\qv}\d_{\rm conv,\qv} \ , 
\label{eqn:G}
\ee
where $\delta$, $\delta_{\rm mask}$ and $\delta_{\rm conv}$ are given in section (\ref{sec:section2}).
Inserting eq. \eqref{eqn:G} into eq. \eqref{eqn:Cmix2a} we end up with four mix terms that we call $C_{\rm mix}^{i}$, with $i=1...4$.
Let's start with the first contribution that come from the first line of eq. \eqref{eqn:Cmix2a}:
\bea
C_{\rm mix}^{1} &=& \frac{1}{N_{k_i}N_{k_j}} \sum_{\qv \in k_i}\sum_{\pv \in k_j}\langle  \d_{\pv}\d_{-\pv}\d_{\rm mask,\qv}\d_{\rm conv,-\qv} - \d_{\pv}\d_{-\pv}\d_{\qv}\d_{\rm conv,-\qv} - \nn \\
& &\d_{\pv}\d_{-\pv}\d_{\qv}\d_{\rm mask,-\qv}\rangle = \nn \\
& & \frac{1}{N_{k_i}N_{k_j}} \sum_{\qv \in k_i}\sum_{\pv \in k_j} \int d^3s \ \langle \d_{\pv}\d_{-\pv}\d_{-\qv} \rangle \langle \d_{\rm mask,\qv}\d_{\rm mask,-\qv}\rangle + \rm{cc}\  = 0 \ ,
\label{eqn:Cmix1}
\eea

\noindent
for all $\qv$ different from zero, with cc for complex conjugate.
For the same reason $C^2_{\rm mix} = 0 = C^3_{\rm mix}$.
The only non zero contribution is:
\bea
C_{\rm mix}^{4} &=& \frac{1}{N_{k_i}N_{k_j}} \sum_{\qv \in k_i}\sum_{\qv \in k_j} [ \langle \d_{\rm mask,\qv}\d_{\rm mask,\pv}\d_{\rm conv,-\qv}\d_{\rm conv,-\pv} -  \nn \\
& & \d_{\rm mask,\qv}\d_{\pv}\d_{\rm conv,-\qv}\d_{\rm conv,-\pv} - \d_{\rm mask,\qv}\d_{\pv}\d_{\rm conv,-\qv}\d_{\rm mask,-\pv} - \nn \\
& & \d_{\qv}\d_{\rm mask,\pv}\d_{\rm conv,-\qv}\d_{\rm conv,-\pv} + \d_{\qv}\d_{\pv}\d_{\rm conv,-\qv}\d_{\rm conv,-\pv} + \nn \\ 
& & \d_{\qv}\d_{\pv}\d_{\rm mask,-\pv}\d_{\rm conv,\qv} - \d_{-\qv}\d_{\rm mask,\qv}\d_{\rm mask,\pv}\d_{\rm conv,-\pv} + \nn \\
& & \d_{\qv}\d_{\rm mask,-\qv}\d_{\pv}\d_{\rm conv,-\pv} + \d_{\qv}\d_{\rm mask,-\qv}\d_{\pv}\d_{\rm mask,-\pv}\rangle ] = \nn \\
& & \frac{1}{N_{k_i}N_{k_j}} \sum_{\qv \in k_i}\sum_{\pv \in k_j} \Bigl\{\int d^3s_1 d^3s_2 \ \langle \d_{-\qv}\d_{-\pv} \rangle \langle \d_{\rm mask, \qv}\d_{\rm mask, \pv}\d_{\rm mask, -\qv+\sv_1}\d_{\rm mask, -\pv+\sv_2} \rangle + \nn \\
& & \qquad \qquad \qquad \quad \int d^3s_1 d^3s_2 \ \langle \d_{\qv}\d_{\pv}\d_{-\qv}\d_{-\pv} \rangle \langle \d_{\rm mask, -\qv+\sv_1}\d_{\rm mask, -\pv+\sv_2} \rangle + \nn \\
& & \qquad \qquad \qquad \quad \langle \d_{\qv}\d_{\pv} \rangle \langle \d_{\rm mask, -\qv}\d_{\rm mask, -\pv} \rangle \Bigr\}
\label{eqn:Cmix4}
\eea 
As we can see from the previous relations, the mixed terms are
convolutions of high order correlators of both cosmological and
mask fields. 

It is possible to further expand the expressions, considering that all
the 4-point correlators can be written in the form: \bea \langle
\d_1\d_2\d_3\d_4\rangle = \langle \d_1\d_2\d_3\d_4\rangle_{\rm
  connected} + \langle \d_1\d_2\rangle\langle\d_3\d_4\rangle +
\rm{perm}. \ .  \eea We expect that the cosmological connected part
are equal to zero, but we can not assume that the same is valid for
the mask connected part.

\section{Realistic galaxy bias}
\label{sec:bias}
In section (\ref{sec:section2}) we made the simplifying assumption
that the quantity $\Phi(\xv;L)$ can be factorized into
luminosity-dependent and position-dependent functions
(equation~\ref{eq:nobias}) This condition simplifies the calculations,
but it is clearly an approximation sincees it implies
luminosity-independent bias. In this section we show how to include a
more realistic bias description. We will then compare the predictions
for the observed power spectra with the ``simplified'' condition. As
in the main text, we will use halo mass as a proxy for luminosity.

Let us consider the number density of halos with mass greater than the nominal threshold $M_0$ to be given by
\be
  n(\xv,M_0) = \int_{M_0}^{\infty} dM \ \Phi(\xv;M) \ , 
\ee
then the observed number density, after the mask perturbation is
\be
  n^{obs}(\xv;M_0) = \int_{M_0\,[1+A(\tv)]}^{\infty} dM \ \Phi(\xv;M)\,.
\ee
For small perturbations to the mass threshold, i.e. $\sigma_A\lesssim 1$, we can Taylor-expand $n^{obs}$ with respect to $A(\tv)$ to get
\bea
n^{obs}(\textbf{x};M_0) & = & n(\textbf{x},M_0) + \frac{\partial n^{obs}}{\partial A}\Bigr|_{A=0}A \nn \\
 & & +  \frac{1}{2}\frac{\partial^2 n^{obs}}{\partial A^2}\Bigr|_{A=0}A^2 + \mathcal{O}(A^3) \ .
  \label{eqn:Ntaylor}
\eea
that we can formally express as
\bea
n^{obs}(\xv,M_0) & = & n(\xv,M_0) +  n^{(1)}(\xv,M_0)A(\tv) \nn\\
& & +  n^{(2)}(\xv,M_0)A^2(\tv) + \mathcal{O}(A^3) \ , 
  \label{eqn:Ntayolr2}
\eea
where
\bea
  n^{(1)}(\xv,M_0) &\equiv& \frac{\partial n}{\partial(\ln M_0)} \\ 
  n^{(2)}(\xv,M_0) &\equiv& \frac{1}{2} \frac{\partial^2 n}{\partial(\ln M_0)^2} \ . 
  \label{eqn:N1N2}
\eea
We can now expand in $A$ the halo density contrast, defined as
\be
  \delta_h^{obs}(\xv,M_0) \equiv \frac{n^{obs}(\xv,M_0)}{\bar{n}^{obs}(\xv,M_0)} - 1 \ ,
\ee
and obtain
\bea
  \delta_h^{obs} & = & \delta_h(1-\frac{1}{2}C_2\sigma_A^2) +  C_1A + (C_1\delta_h + \tilde{\epsilon})A \nn\\
  & & + \frac{1}{2}C_2(A^2-\sigma_A^2) +\frac{1}{2}(C_2\delta_h+2C_1\tilde{\epsilon}+\tilde{\eta})A^2 \nn \\ 
  & & + \mathcal{O}(A^3) \ .
  \label{eqn:dobs_conf}
\eea
where $\delta_h$ is the cosmological halo density contrast and where we defined
\bea
  C_1(M_0) & \equiv & \frac{\partial \ln \bar{n}}{\partial \ln M_0} \\
  C_2(M_0) & \equiv & \frac{M_0^2}{\bar{n}}\frac{\partial^2 \bar{n}}{\partial M_0^2} \\
  \tilde{\epsilon} & \equiv & M_0\frac{\partial \delta_h}{\partial M_0} \\
  \tilde{\eta} & \equiv & M_0^2\frac{\partial^2 \delta_h}{\partial M_0^2} \ .
\eea
It is easy to see that if eq. \eqref{eq:nobias} holds,
then $\tilde{\eta}=\tilde{\epsilon}=0$ and the density contrast
reduces to the form of eq. \eqref{eqn:delta}. From equation~\ref{eqn:dobs_conf}, one can show that the two-point correlation
function in Fourier space is given by
\bea
  \langle
  \delta_{\kv_1}^{obs}\delta_{\kv_2}^{obs}\rangle & = &
  \delta_D(\kv_{12})P_{\delta\delta}(k_1)(1-C_2\sigma^2_A) +\langle A_{\kv_1}A_{\textbf{k}_2}\rangle \nn \\
  & & + \int d^3 q
     [P_{\delta\delta}(q)+2C_1P_{\delta\tilde{\epsilon}}(q)+C^2_1P_{\tilde{\epsilon}\tilde{\epsilon}}(q)] \nn \\
     & & \times \langle A_{\kv_1-\qv}A_{\kv_1+\qv}\rangle
     \nn \\ 
  & & + \frac{1}{2}
            [C_2P_{\delta\delta}(k_1)+2C_1P_{\delta\tilde{\epsilon}}(k_1)+C_1^3P_{\tilde{\epsilon}\tilde{\eta}}(k_1)] \nn \\
& & \times \int d^3 q \langle  A_{\qv}A_{\kv_{12}-\qv}\rangle \nn \\
            &  &+(\kv_1 \leftrightarrow \kv_1) 
  + \mathcal{O}(A^3) \ .
\eea
Since we are interested in a comparison with eq. \eqref{eqn:PobsEstim}, using eq. \eqref{eqn:deltaObsK} with $\delta^{obs}$ from eq. \eqref{eqn:dobs_conf} we obtain the monopole of the observed power spectrum to be given by
\bea
P_0^{obs} & = & P_{\delta\delta}(1-C_2\sigma_A^2) + P_{AA,0}(k)
  \nn \\ 
  & & +\int d^3 q
         [P_{\delta\delta}(q)+2C_1P_{\delta\tilde{\epsilon}}(q)+C^2_1P_{\tilde{\epsilon}\tilde{\epsilon}}(q)]P_{AA,0}(|\qv-\kv|)
         \nn  \\ 
 &  & +[C_2P_{\delta\delta}(k_1)+2C_1P_{\delta\tilde{\epsilon}}(k_1)+C_1^3P_{\tilde{\epsilon}\tilde{\eta}}(k_1)]\sigma_A^2 
 \nn \\
& & + \mathcal{O}(A^3) \ ,
\eea
where
\be
 P_{AA,0}(k)= \frac{k_f^3}{V_p(k)} \int_kd^3\textbf{q}\langle|A_{\textbf{q}}|^2\rangle
\ee
is the monopole of the mask power spectrum. The main difference with
\eqref{eqn:PobsEstim}, is given by the presence of the power spectra $P_{\delta\tilde{\epsilon}}$, $P_{\delta\tilde{\eta}}$ and
$P_{\tilde{\eta}\tilde{\eta}}$, all dependent on derivatives of the density contrast $\delta_h$. Considering a local bias expansion as
$ \delta_h = \sum_n b_n(M)\delta_m^n $ these additional terms could be expressed in terms of  derivatives of the halo bias functions $b_n(M)$.

A modulation of the mass threshold implies a different halo selection,
and then a different bias. Because the relation between bias and
threshold halo mass is not linear, we expect that the bias of halos
subject to a mask $A(\tv)$ will be different from the bias
of unmasked halos, even when $\langle A(\tv) \rangle=0$. As a
consequence, we expect the masked and unmasked catalogs not to match
at small scales, as they do in Figure~\ref{fig:PS_A5A20}.

\begin{figure}
  \centering
  \includegraphics[width=0.9\textwidth]{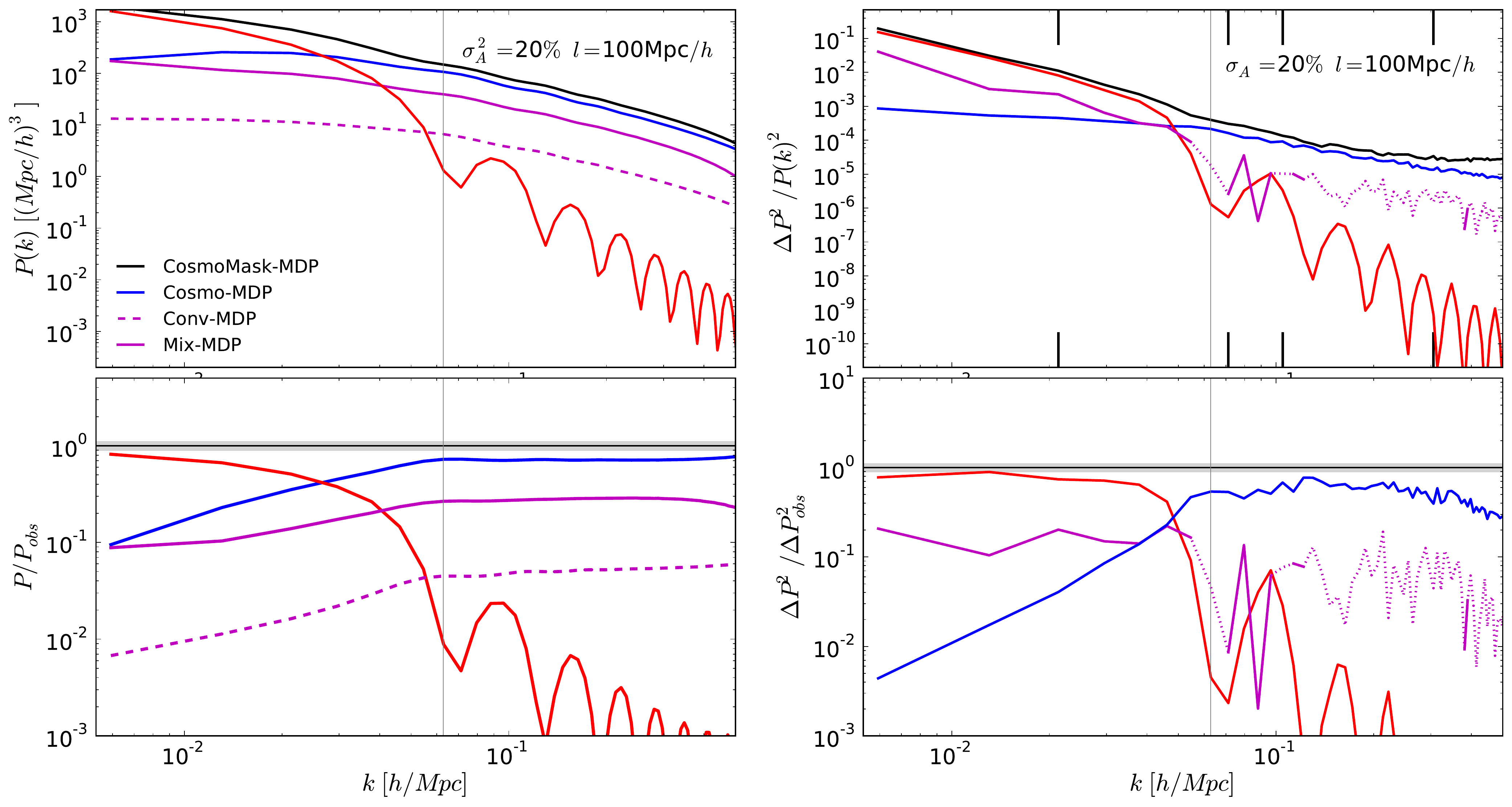}
  \caption{Averaged power spectra and variance components without mass
    bias removal. Color code is the same of figures
    (\ref{fig:PS_A5A20}, \ref{fig:Var_A5A20}).}
  \label{fig:PS_bias}
\end{figure}

We will not go into the computation details of the full covariance
matrix, because we can directly use mocks to see the effect of
introducing the halo bias (fig.~(\ref{fig:PS_bias})). In this figure
we show all the power spectrum and variance components; ``MDP''(Mass
Dependent Bias) marks the terms obtained by using the original DM halo
masses, without removing the mass dependency of bias. As expected, at small
scales we observe a difference of few percent between the
cosmological and the  masked power spectrum, due to the mass
bias. The same effect is observable in the power spectrum variance at
the same scales. Of course this makes the estimation of the
convolution term of the power spectrum, and the of mixed terms in the
covariance matrix, more complicated. In the figure we show again as
magenta lines these terms computed by difference as in the main text,
but the large difference between the measurement of the convolution
term in the power spectrum and its theoretical prediction shows that
there is a problem.

\acknowledgments

This work has been triggered by a discussion, within the Euclid
Consortium, on how to model the total covariance of 2-point
clustering. We thank Enzo Branchini for carefully reading this paper,
and Michele Maris for discussions. We acknowledge financial support
from PRIN-MIUR 201278X4FL, PRIN-INAF 2012 “The Universe in a Box”, the
INFN INDARK grant and “Consorzio per la Fisica” of Trieste.
Simulations have been carried out thanks to a CINECA-UNITS agreement.
Data have been stored on the CINECA facility PICO, granted us thanks
to our expression of interest.



\bibliographystyle{JHEP}
\bibliography{biblio}



\end{document}